\documentclass [11pt,a4paper]{article}
\usepackage{jheppub}
\usepackage[utf8]{inputenc}
\usepackage{graphicx}
\usepackage{color}
\usepackage{amsfonts,amsthm,amsmath,amssymb}
\newcommand\mb{\, \overset{\mathrm{\star}}{,} \, }
\usepackage{comment}

\usepackage{graphicx}
\usepackage{color}
\usepackage{bm,bbm}
\usepackage[utf8]{inputenc}
\usepackage{graphicx,color}
\usepackage{amsfonts,amsthm,amssymb}
\usepackage[english]{babel}
\usepackage{verbatim}
\usepackage{mathtools}

\newcommand{\be}{\begin{equation}}
\newcommand{\ee}{\end{equation}}
\newcommand{\ba}{\begin{eqnarray}}
\newcommand{\ea}{\end{eqnarray}}
\def\bs{\begin{subequations}}
\def\es{\end{subequations}}

\def\com{\color{magenta}}
\def\cob{\color{blue}}

\newcommand{\book}[5]{\emph{#1} (#2, #3, #4, #5)}

\newcommand{\oarX}[1]{\href{http://arxiv.org/abs/#1}{{\ttfamily\com arXiv:#1}}}
\newcommand{\arX}[1]{\href{http://arxiv.org/abs/#1}{{\ttfamily\com arXiv:#1}}}
\newcommand{\doin}[6]{\href{http://dx.doi.org/#1}{{\cob #2 #3 {\bf #4}, #5 (#6)}}}
\newcommand{\doinn}[5]{\href{http://dx.doi.org/#1}{{\cob #2 {\bf #3}, #4 (#5)}}}

\newcommand{\tia}[1]{\textit{#1},}

\newcommand{\bea}{\begin{eqnarray}}
\newcommand{\eea}{\end{eqnarray}}

\def\lp{\ell_{\rm Pl}}

\def\mpl{m_{\rm Pl}}

\begin{document}

\title{Extending general covariance:\\ Moyal-type noncommutative manifolds}

\author{Martin Bojowald$^{a,*}$,}
\emailAdd{bojowald@gravity.psu.edu$^*$}
\affiliation{$^{a}$Institute for Gravitation and the Cosmos, The Pennsylvania State University,104 Davey Lab, University Park, PA 16802, USA}

\author{Suddhasattwa Brahma$^{a,b,c,\dagger}$,}
\emailAdd{suddhasattwa.brahma@gmail.com$^{\dagger}$}
\affiliation{$^{b}$Center for Field Theory and Particle Physics, Fudan University, 200433 Shanghai, China}
\affiliation{$^{c}$Asia Pacific Center for Theoretical Physics, Pohang 37673, Republic of Korea}

\author{Umut Buyukcam$^{a,\dagger\dagger}$, and}
\emailAdd{uxb101@psu.edu$^{\dagger\dagger}$}

\author{Michele Ronco$^{a,d,e,**}$}
\emailAdd{michele.ronco@roma1.infn.it$^{**}$}
\affiliation{$^{d}$Dipartimento di Fisica, Universit\`a di Roma ``La Sapienza,'' P.le A.\ Moro 2, 00185 Roma, Italy}
\affiliation{$^{e}$INFN, Sez.~Roma1, P.le A.\ Moro 2, 00185 Roma, Italy}

\date{September 20, 2017}

\abstract{In the Hamiltonian formulation of general relativity, Einstein's
  equation is replaced by a set of four constraints. Classically, the
  constraints can be identified with the generators of the
  hypersurface-deformation Lie algebroid (HDA) that belongs to the groupoid of
  finite evolutions in space-time.  Taken over to deformed general relativity,
  this connection allows one to study possible Drinfeld twists of space-time
  diffeomorphisms with Hopf-algebra techniques. After a review of
  noncommutative differential structures, two cases --- twisted
  diffeomorphisms with standard action and deformed (or $\star$-)
  diffeomorphisms with deformed action --- are considered in this paper. The
  HDA of twisted diffeomorphisms agrees with the classical one, while the HDA
  obtained from deformed diffeomorphisms is modified due to the explicit
  presence of $\star$-products in the brackets. The results allow one to
  distinguish between twisted and deformed symmetries, and they indicate that
  the latter should be regarded as the relevant symmetry transformations for
  noncommutative manifolds. The algebroid brackets maintain the same general
  structure regardless of space-time noncommutativity, but
    they still show important consequences of non-locality. }

\maketitle

\section{Introduction}

Thanks to general relativity (GR), gravitational interactions are understood
as purely geometric phenomena which can be described in terms of a metric, an
affine connection, and a curvature defined on a (pseudo-)Riemannian
manifold. The symmetry of general covariance is an important governing
principle which determines possible dynamical theories. Accordingly, one may
attempt to quantize gravity by analyzing possible quantum space-time
symmetries which determine the structure of the geometry of the system. As
shown in
Refs.~\cite{dfr1,dfr2,ven,padma,kon,ellis,gara,alu,ngdam,amelino,kempf,maggiore,frafuzz1,frafuzz2},
the concept of absolutely sharp points, one of the cornerstones of Riemannian
geometry, should then be expected to break down. A general mathematical
structure that can make sense of such a space-time is provided by
noncommutative geometry
\cite{conn1,conn2,szabo1,szabo2,douglas2,madore,douglas1,posop} which involves
a notion of deformed symmetries often referred to as quantum groups
\cite{qg1,qg2}. For more than twenty years now, the study of possible quantum
deformations of relativistic symmetries has been intensely pursued
\cite{qg3,qg4,qg5,qg6,qg7,qg8,qg9,qg10,qg11}, and supersymmetric extensions
have been considered as well \cite{qss1,qss2,qss3}. The present understanding
is that, even with noncommutativity, it is still possible to have a
ten-dimensional local symmetry group (replacing classical Poincar\'e
transformations) by means of what is known as a `Drinfeld twist'
\cite{drin,DT,cattaneo}.

In the case of flat space-time, twists allow one to interpret noncommutative
versions of Minkowski spacetime as objects which are, in a certain sense, dual
to suitable deformations of the Poincaré algebra. Identifying the dimensionful
deformation parameter $\lambda$ (or $\kappa \sim 1/\lambda$) with the Planck
length $\lp = \sqrt{\hbar G/ c^3}$ (Planck mass $\mpl = \sqrt{c\hbar/G}$),
these models provide a mathematical realization of the proposal of doubly (or
deformed) special relativity \cite{dsr1,dsr2}, which argues that Planck-scale
effects should necessitate a description of space-time physics in terms of two
relativistic invariants: $\lambda$ or $\kappa$ in addition to the speed of
light $c$. In spite of this success, the extension of noncommutativity to
curved manifolds remains an open issue, which is of particular importance
because one of the main applications of quantum groups and the associated
space-time noncommutativity is the characterization of Planck-scale
physics. They should therefore have the potential to be a candidate theory of
quantum gravity (QG), which has to include curved space-time solutions.

Attempts to quantize 3-dimensional gravity have met with more success
\cite{3d1,3d2,3d3,3d4,3d5,3d6,3d7,3d8,3d9,3d10}, but much work remains to be
done to generalize these results to the 4-dimensional theory of physical
interest. Understanding how to quantize GR or, even more generally, the class
of all possible covariant theories remains center stage in the research
program of noncommutative geometry and the associated deformation of gauge
groups. Another important stimulus to study the deformation of diffeomorphisms
groups, which can be regarded as the gauge symmetries of GR, comes from string
theory. In this context, it has been shown that coordinates obey canonical
noncommutativity if a background tensor field (or $B$-field in short) is
present \cite{witten1,witten2,witten3,sst,boer}. 

In the last two decades, the study of Hopf algebras from a physical
perspective has given rise to a rather sizable literature on quantum Minkowski
spacetimes \cite{qm1,qm2,qm3,qm4}. These zero-curvature models are often
considered toy models for the flat limit of a (still to be found) QG
theory. In some very rare cases, they have even proved useful for
phenomenological proposals \cite{gacLRR}. The main idea is to promote
coordinates $x^\mu$ to noncommuting operators $\widehat{X}^\mu$ with
non-trivial commutators of the form $[\widehat{X}^\mu, \widehat{X}^\nu] = i
\theta^{\mu\nu}(\widehat{X}) = i\theta^{\mu\nu} + i \Theta^{\mu\nu}_\rho
\widehat{X}^\rho$. Thanks to Weyl-Moyal maps, which had been first introduced
to study the phase space of quantum mechanics, one can trade operator-valued
coordinates for functions living on a classical manifold but equipped with a
non-standard multiplication rule. This procedure introduces a noncommutative
$\star$-product, whose main feature is non-locality. Such quantum deformations
of coordinate spaces based on algebraic relations have been extensively
studied since the seminal paper by Snyder \cite{snyder}. The best known
examples are given by $\theta$ (or Moyal) canonical space \cite{can},
$\kappa$-Minkowski spacetime invariant under the $\kappa$-Poincar\'e algebra
\cite{majrue,lukrue}, $q$-deformations of Lie groups \cite{qp1,qp2,qp3}, and
the fuzzy sphere \cite{madore,bala}.

All this literature mainly focused on the construction of noncommutative
Minkowski space-times but did not contemplate extensions to curved
versions. Some progress has been made in the quantization of symmetry-reduced
GR solutions such as DeSitter \cite{ds}, anti-DeSitter \cite{ads}, and FRW
\cite{frw} backgrounds. Nevertheless, the situation for the quantization of
the full group of diffeomorphisms remains unclear and the relevant literature
is fragmented. The main obstacle seems to be the proper definition of
coordinate transformations and a self-consistent calculus once coordinates
have been promoted to noncommuting objects. It is not difficult to realize
that noncommutativity introduces a preferred frame (or coordinate choice) and
thus is not compatible with the standard symmetries. For instance, if we
assume that $[\widehat{X}_\rho, \widehat{X}_\sigma] = i \theta_{\rho \sigma}$,
as it is the case for the canonical or Moyal-Weyl noncommutative spacetime,
then the transformed coordinates $\widehat{X'}_\mu = \widehat{X}_\mu +
\widehat{\xi}_\mu$, with a vector field $\widehat{\xi}_\mu$ depending linearly
on $\widehat{X}_\mu$ (as required for rotations and boosts), do not obey the
original commutation relation $[\widehat{X'}_\rho, \widehat{X'}_\sigma] \neq i
\theta_{\rho \sigma}$. To avoid this, as we briefly hinted above, one needs to
quantize (or deform) the symmetry group in a specific way. Such a deformation
theory in complete form is not available for diffeomorphism groups. For this
reason, we do not yet have a widely accepted noncommutative theory of gravity.

A possible way out, proposed in Ref.~\cite{calmet}, lies in restricting the
group of diffeomorphisms to those transformations that preserve coordinate
noncommutativity. It has been recognized \cite{calmet} that, in the case of
canonical space, this proposal corresponds to a restriction to
volume-preserving diffeomorphisms. One therefore obtains a connection with
unimodular gravity \cite{unim1,unim2}. Another possibility, explored in
Ref.~\cite{cham}, is a generalization of the Seiberg-Witten map \cite{witten1}
to GR by gauging the Lorentz group. A drawback of such an approach is that it
forces one to use a complex metric structure \cite{cham}. An alternative
perspective on the interplay between gravity and noncommutative geometry is
offered for instance by Refs.~\cite{har1,har2}.

Perhaps one of the most promising paths proposed so far is that of twisted
diffeomorphisms \cite{aschieri1,aschieri2}. The main idea is to replace the
diffeomorphism invariance of GR by its twisted version. This is done by
deforming the Hopf algebra structure of the universal enveloping algebra of
the Lie algebra of vector fields by twisting the coproduct by means of
Drinfeld twists \cite{aschieri1,aschieri2}. The action of diffeomorphisms on
single fields then stays unmodified while the Leibniz rule (which provides the
action on two or more fields) is changed. As a result, the $\star$-product of
two (or more) fields is covariant under twisted diffeomorphisms. Finally, one
can write down a modification of the Einstein-Hilbert action which is
invariant under twisted diffeomorphisms thanks to an appropriate
$\star$-product. Given the potential of such an approach, Ref.~\cite{gaume}
explored whether such a (twisted) noncommutative gravity can be obtained from
closed strings with an external $B$-field in the Seiberg-Witten
limit. Unfortunately, there has been no way of matching this limit of string
theory with the gravity model of Ref.~\cite{aschieri1}. Moreover, as already
pointed out in Refs.~\cite{gaume,mozo}, we stress that twisted symmetries are
not genuine deformations of classical symmetries but rather mappings of the
classical symmetries on spaces with noncommutative $\star$-products. Following
what has been done for other gauge groups \cite{szabo1,szabo2,douglas2}, one
should properly deform also the action on single fields in order to have a
definition of $\star$ (or deformed) diffeomorphisms. To our knowledge, no
such formulation is currently available in the literature.  The introduction
of deformed diffeomorphisms, as opposed to twisted diffeomorphisms, represents
one of the main objectives of the present work.

In addition, we propose a new line of inquiry and ask whether diffeomorphisms
can be consistently quantized in the sense of a deformation theory in analogy
to what has been already done for the special relativistic (SR) group of
Poincar\'e symmetries. We therefore provide candidate structures for any
deformed general relativistic theory, without using specific actions or
dynamical equations.  In contrast to most previous studies of noncommutative
geometry, we follow a canonical approach.  Along the lines of the classical
analysis of Dirac \cite{dirac} and Arnowitt-Deser-Misner (ADM) \cite{adm}, it
should be possible to perform a $3+1$-splitting of the action of
Ref.~\cite{aschieri1}. Poisson brackets of the resulting scalar and momentum
constraints would then lead to the corresponding hypersurface-deformation
algebroid (HDA) or Dirac spacetime algebra
\cite{dirac,teit,bojo1}. Unfortunately, however, the full ADM machinery turns
out to be rather involved when it is applied to gravitational actions on
noncommutative manifolds.

As we point out in this paper, there is a shortcut that can provide us with
general (that is, action-independent) hints for hypersurface deformations or
diffeomorphisms on such manifolds. It therefore leads us to a notion of
deformed general covariance. The shortcut is motivated by recent results
of Ref.~\cite{wein} for classical smooth manifolds, further developed in
Ref.~\cite{bojo2} under weaker assumptions that allow one to bring in some
quantum-gravity effects. For our purposes here, the main achievement is the
recognition that the symmetry structure of hypersurface deformations (which is
usually described as a ``Lie algebra with structure functions'' in the
physics literature) is that of a Lie algebroid which can be derived from a
groupoid of finite evolutions between space-like hypersurfaces in Lorentzian
manifolds. (A similar Euclidean version also exists.) In particular, the
rather complicated Poisson brackets between the gravitational constraints of
canonical gravity are reproduced by the tangential and normal components of
Lie brackets between suitable (Gaussian) space-time vector fields.  In order
to inspect the HDA for noncommutative spacetimes, it is then not necessary to
know the explicit expressions of constraints as phase-space functions, which
in fact would not be available for noncommutative gravity. It is sufficient to
introduce a suitable differential calculus and apply it to such a
noncommutative version of a tangential-normal decomposition by following the
steps of recent analyses \cite{wein,bojo2}, observing certain consistency
conditions extracted from \cite{wein}.

We will start by modifying general coordinate transformations of commuting
variables into diffeomorphisms of noncommuting functions. Moyal-Weyl maps
allow us to treat operator-valued objects as standard functions, but
multiplied with a noncommutative $\star$-product.  (That is, to introduce
noncommutativity we do not need to change the classical function space, but
only the product in the algebra of functions.)  At the classical level,
infinitesimal diffeomorphisms form a Lie algebra with an extension of their
action from vector fields to tensor fields because the standard Leibniz rule
applies. We deform this structure by using Drinfeld twists and, thus, define a
deformed differential geometry. When we analyze the case of twisted
diffeomorphisms, the algebra remains unchanged while the comultiplication
changes, confirming the suggestions made in
Refs.~\cite{aschieri1,aschieri2}. Twisted diffeomorphisms are opposed to
deformed (or $\star$-) diffeomorphisms which we introduce and discuss for the
first time. 

In the definition of the action on single fields we follow established results
in the literature, while we explore two possibilities regarding the
comultiplication rule or coalgebra sector of $\star$-diffeomorphisms. We will
first try to mimic the situation of $U(N)$ noncommutative gauge theories
\cite{szabo1,douglas1,douglas2,ncqft1,ncqft2,ncqft3,ncqft4,ncqft5} and work
with trivial coproducts. (The standard Leibniz rule then applies.) We will
note several drawbacks of retaining the standard Leibniz rule, which leads us
to propose a suitable deformation of comultiplication. In both cases we are
able to compute the HDA brackets and show that, as opposed to the twisted
case, there are $\star$-product deformations in the algebra which distinguish
deformed from twisted diffeomorphisms. Sharing the concerns raised in
Ref.~\cite{gaume}, we expect that $\star$-diffeomorphisms, rather than twisted
ones, should be chosen as the symmetries of a noncommutative theory of
gravitation.

Together with previously established results in the literature on
noncommutative gravity, our work provides general results about possible
formulations of a deformed gravity theory, defined with a deformed
differential geometry on noncommutative hypersurfaces. The closed
brackets of hypersurface-deformation generators with star products found
here can be used to test the covariance of existing proposals for
noncommutative gravity theories, but they may also prove useful in the
construction of new such models or in a classification of all possible
deformations of classically covariant theories.

To some extent, noncommutative gravity represents an independent approach to
QG.  However, we wish to stress that, besides the aforementioned seminal
papers \cite{witten1,witten2,witten3} showing the appearance of
noncommutativity in string theory due to the presence of external fields,
additional support to a possible role for spacetime noncommutativity in string
theory has been recently claimed in
Refs.~\cite{freidelST1,freidelST2,freidelST3}: There, it has been shown that
the target space of closed strings is noncommutative regardless of the
specific features of the background. Additional motivation for our work comes
from the recent interest in modifications and/or generalizations of the HDA
found in the QG literature
\cite{holocorr1,holocorr2,holocorr3,perez,anomlqc,vara,tomlin,frahda,frarev},
including a possible way to ascribe Minkowski spacetime quantization and
Poincar\'e symmetry deformation to loop quantum gravity corrections
\cite{paily,nclqg,roscian,dani,phenolqg,phenolqg1,loopdim1,loopdim2}. General
deformations of the HDA have also been studied recently in Ref.~\cite{vasu},
where the authors found a (partial) no-go theorem forbidding specific
modifications of the scalar constraint in a general covariant theory. It is
possible to regard our work as an explicit example showing that the
assumptions of such a theorem can be weakened so as to evade the original
conclusions.

Our paper is organized as follows. In Section~\ref{s:Math}, we first review
the definitions of Lie algebroids and rederive the classical HDA starting from
the Lie brackets of a suitable class of space-time vector fields. Then, we
remind the reader of the notions of Hopf algebras and introduce a differential
calculus on noncommutative manifolds. Vectors, differential forms, tensors,
$\star$-Lie derivatives, inner products, and index contraction are all
defined. Two different notions of brackets are introduced --- Moyal and
$\star$-Lie brackets --- together with a discussion of their
differences. Section~\ref{s:Twisted} is dedicated to the analysis of
hypersurface deformations generated by twisted diffeomorphisms with the Moyal
$\star$-product. After defining a proper modification of the classical
condition on space-time vector fields, we compute the Lie brackets between
them and then decompose the result into normal and tangential parts, thereby
obtaining a twisted version of the HDA. Confirming the expectations of
Refs.~\cite{aschieri1,aschieri2}, we find that the HDA is unmodified. This
result also ensures that twisted gravity possesses the same degrees of freedom
as classical GR. In Section~\ref{s:Deformed}, we focus on deformed
diffeomorphisms. Two different possibilities for the coalgebra sector are
considered before an analysis analogous to the previous case is carried
out. The resulting HDA is deformed due to the presence of explicit
$\star$-product contributions. Finally, we draw our conclusions and sketch an
outlook in Section~\ref{s:Concl}.


\section{Mathematical preliminaries}
\label{s:Math}

The main mathematical tools used here, Lie algebroids and Hopf
algebras, are reviewed in this section.

\subsection{Lie algebroids}

We closely follow \cite{wein,bojo2} but similar content can also be found, for
instance, in \cite{xu}.  A Lie algebroid is a vector bundle $A$ over a smooth
base manifold $B$ together with a Lie bracket $[\cdot ,\cdot ]_A$ on the set
$\Gamma(A)$ of sections of $A$ and a bundle map $\rho\colon \Gamma(A)
\rightarrow \Gamma(TB)$, called the anchor, provided that the following two
properties are satisfied:
\begin{itemize}
\item $\rho\colon ( \Gamma(A), [\cdot ,\cdot ]_A) \rightarrow \left(\Gamma(TB),
    [\cdot ,\cdot]\right)$ is a Lie-algebra homomorphism: for any $\xi, \eta
  \, \in \, \Gamma(A)$, we have $\rho([\xi, \eta]_A) = [\rho(\xi),
  \rho(\eta)]$ (the Lie bracket of vector fields in $\Gamma(TB)$).
\item For any $\xi, \eta \in \Gamma(A)$ and $f \in C^\infty (B)$, the Leibniz
  rule $[\xi,f\eta]_A =f[\xi,\eta]_A+(\rho(\xi)f)\eta$ holds.
\end{itemize}
If the base manifold $B$ is a point, the Lie algebroid is a Lie algebra. Let
us also mention that, in the case of Lie algebroids, one needs to generalize
the notion of Lie algebra morphisms if one desires to identify classes of
equivalence. However, morphisms between algebroids will not play any role in
our analysis. We refer the interested reader to Ref. \cite{xu} and references
therein.

We are primarily interested in the specific Lie algebroid of
hypersurface-deformations, which provides a mathematical formulation of the
Poisson brackets of gravitational constraints \cite{dirac,teit,adm}.  Gauge
transformations generated by the constraints are equivalent to space-time
diffeomorphisms. In a canonical formulation, invariance under these
transformations ensures that observables of the theory are independent of the
particular embedding of spatial hypersurfaces in space-time. An explicit
derivation of hypersurface-deformation brackets can make use of coordinate
choices to simplify calculations. The closure of the brackets in the form of a
Lie algebroid then ensures that they are covariant under changes of the
embedding.

A convenient choice turns out to be given by Gaussian embeddings,
which are defined such that the space-time metric $g_{\mu\nu}$ assumes a
Gaussian form:
\begin{equation}
{\rm d}s^2 = -{\rm d}t^2 + q_{ab} {\rm d}x^a {\rm d}x^b \, .
\end{equation}
Thus, for the components of $g_{\mu\nu}$ one has
\begin{equation}\label{cgaus}
  g_{\mu\nu} = - n_\mu n_\nu + q_{ab}X^a_\mu X^b_{\nu} 
\end{equation}
with the spatial metric $q_{ab}$.  We have written the metric in a basis dual
to $(n^\mu, X^\mu_a)$, where $n^\mu$ is the unit normal to a family of
space-like hypersurfaces $\Sigma_t$ (at constant $t$), while $X^\mu_a$ form a
basis of $T\Sigma_t$. With these conditions, we have the orthonormality
relations $g_{\mu\nu}n^\mu n^\nu = -1$ and $g_{\mu\nu}n^\mu X^\nu_a =
0$. Following the ADM treatment of canonical gravity \cite{adm}, we then
decompose the time-evolution vector field $\tau^\mu$ by $\tau^\mu = N n^\mu +
M^a X^\mu_a$, where $N$ is the lapse function and $M^a$ the shift vector
field.

A foliation which is Gaussian for one embedding is, in general, not Gaussian
for a different embedding. Gaussianity is therefore not preserved by general
coordinate transformations. We can, however, restrict the class of
transformations to diffeomorphisms generated by {\em Gaussian} vector fields
$v^\mu$ obeying
\begin{equation}
i_n \mathcal{L}_v g = 0 \, , 
\end{equation}
or, in components, 
\begin{equation}
n^\mu \mathcal{L}_v g_{\mu\nu} = 0 \, .
\end{equation}
Here (and throughout the paper) $i_w$ stands for the internal product (or
contraction) with a vector field $w$.  The normal components of the metric
remain invariant under transformations along the direction of such a $v^\mu$,
preserving the Gaussian form.  Choosing a Gaussian embedding corresponds to
fixing a representative in each equivalence class of hypersurface embeddings,
in which the subset of Gaussian $v^\mu$ furnishes the remaining coordinate
freedom.

Expanding the Lie derivative, the Gaussian condition can be rewritten as
\begin{equation}
n^\mu v^\rho \partial_\rho g_{\mu\nu} + n^\mu (\partial_\mu v^\rho) g_{\rho \nu}
+ n^\mu (\partial_\nu v^\rho) g_{\rho \mu}  = 0 \, , 
\end{equation}
resulting in
\begin{equation}
v^\rho {\rm d}n_{\rho \nu} + \partial_\nu(v^\rho  g_{\rho \mu} n^\mu) +  g_{\mu\nu}
[n,v]^\mu = 0 \, . 
\end{equation}
We used the Cartan identity, the definition of the Lie bracket, and 
$({\rm d} n)_{\mu\nu} = \partial_\mu n_\nu - \partial_\nu n_\mu$. Due to
the Gaussian from of the metric \eqref{cgaus}, we have ${\rm d}n = 0$
because $n={\rm d}t$ is closed. Decomposing the Gaussian vector in the
basis chosen above --- that is, writing $v^\mu = N n^\mu + M^a X^\mu_a$ --- we
then have 
\begin{equation}
-\partial_\nu N +  g_{\mu\nu}(n^{\mu}n^{\rho}\partial_{\rho}N+ [n, M]^\mu) = 0 \, ,
\end{equation}
where we have used the orthogonality of the basis. (Although we use the same
notation for components $N$ and $M^a$ of a Gaussian vector field and the
time-evolution vector field, the former are more general since they refer
to a coordinate change.)  Projecting this
expression along normal and tangential directions, respectively, we find
\begin{equation}\label{ccomp}
\partial_\nu N = 0 \,  \quad  \mbox{and} \quad \,  [n, M]^a =
q^{ab} \partial_b N \, . 
\end{equation}
Here, $q^{ab}$ is the inverse of the spatial metric. (The bracket
$[n,M]^{\mu}$ does not have a normal component thanks to the geodesic property
of $n^{\mu}$ for a Gaussian system; see \cite{bojo2} for details.)

We can now compute the HDA by calculating the Lie bracket between two
Gaussian vector fields:
\begin{equation} \label{vv}
\begin{split}
[v_1 , v_2 ]^\mu = v^\rho_1 \partial_\rho v^\mu_2 -v^\rho_2 \partial_\rho
v^\mu_1 = (N_1 \mathcal{L}_n N_2 - N_2 \mathcal{L}_n N_1 + \mathcal{L}_{M_1 }
N_2 - \mathcal{L}_{M_2}N_1 )  n^\mu \\ 
+ [M_1 , M_2 ]^\mu + N_1 [n,M_2]^\mu - N_2 [n , M_1]^\mu \\
= (\mathcal{L}_{M_1 } N_2 - \mathcal{L}_{M_2}N_1 )  n^\mu 
+ [M_1 , M_2 ]^\mu + q^{\mu b}(N_1\partial_b N_2-N_2\partial_bN_1)
\, ,
\end{split}
\end{equation}
where we decomposed both $v_1$ and $v_2$ in the basis $(n,X)$, and then used
the equations (\ref{ccomp}). The terms of the type $\mathcal{L}_n N =
n^\rho\partial_\rho N$ are all zero due to the first equality in
\eqref{ccomp}. In order to obtain the HDA, we have to extract normal and
tangential contributions: If $N_1=N_2=0$,
\begin{equation}
 [v_1,v_2]^{\mu} = [M_1,M_2]^{\mu}\,,
\end{equation}
if $M_1^a=0$ and $N_2=0$,
\begin{equation}
 [v_1,v_2]^{\mu} = -n^{\mu}{\cal L}_{M_2} N_1\,,
\end{equation}
and if $M_1^a=0=M_2^a$,
\begin{equation}
 [v_1,v_2]^{\mu} = q^{\mu b}(N_1\partial_bN_2-N_2\partial_bN_1)\,.
\end{equation}
Finally, we view the pairs $(N,M^a)$ as fibers of a Lie algebroid over the
space of spatial metrics, and interpret the three cases of $[v_1,v_2]^{\mu}$
as Lie-algebroid brackets
\begin{eqnarray}\label{classHDA}
&&[(0,M^a_1), (0,M^b_2)] = (0, \mathcal{L}_{M_1}M_2) \, , \\
&&[(N,0) , (0,M^a)]  = \left( -\mathcal{L}_{M} N ,0 \right) \, , \\
&&[(N_1,0) , (N_2,0)] = ( 0,   (N_1  \partial_b N_2 -N_2  \partial_b
N_1)q^{ab} ) \, . 
\end{eqnarray}
(The anchor map is given by the Lie derivative of the metric along
$\tau^{\mu}=Nn^{\mu}+M^a X_a^{\mu}$; see \cite{wein}.)  With these brackets,
pairs $(N,M^a)$ form the hypersurface-deformation Lie algebroid over the space
of spatial metrics. Spatial diffeomorphisms form a subalgebroid which is also
a Lie algebra, while the brackets involving only normal deformations depend on
the inverse-metric components as coordinates on the base manifold (the
``structure functions''). (We note that the base manifold can be extended to
the full phase space of general relativity, given by spatial metrics and
extrinsic curvature, or linear combinations of the latter components. While
this extension is not necessary in the classical algebroid, it may be required
for some quantum effects as we will see later in this paper.)

The derivation presented here has several advantages over the usual ones in
canonical gravity. It is much shorter and minimizes the amount of technical
calculations. Moreover, it utilizes space-time tensor calculus and implements
the $3+1$-split only by decomposing vector fields. It is therefore ideal for
an application to non-classical space-time structures in which some versions
of tensor calculus exist.  The rest of our work is dedicated to an application
of these methods to the deformation theory of this algebroid in order to have
a notion of (deformed) general covariance for noncommutative manifolds. 
  We will focus on the brackets and not discuss the anchor. As shown in
  \cite{xu}, the latter is not subject to deformations.

One question to be discussed in more detail is the definition of Gaussian
systems in non-classical space-times. The Gaussian nature, by itself, is not
relevant because it just constitutes a choice of gauge fixing. However, the
Gaussian system simplifies the classical derivation, and it makes it easier to
check two important consistency conditions which we emphasize here: (i) The
derivation of the hypersurface-deformation brackets requires us to extend the
fields $N$ and $M^a$ from a given hypersurface into a space-time
neighborhood. Only such an extension makes it possible to compute the
space-time Lie derivative of two vector fields in (\ref{vv}) and then
decompose the result into normal and spatial components. In the classical
derivation, such an extension is possible thanks to the form of the
differential equations (\ref{ccomp}), which are well-posed with $N$ and $M^a$
as initial conditions on one hypersurface. (ii) The resulting
hypersurface-deformation brackets (\ref{classHDA}) depend only on spatial
data, given by the fields $N$ and $M^a$ together with the spatial metric
$q_{ab}$. It is therefore possible to interpret them as Lie-algebroid
relations over the space of metrics. There is no dependence on properties of
the embedding of a hypersurface in space-time. 

In our new derivations below, we will take a pragmatic approach and look for a
generalization of the Gaussian condition such that these two consistency
conditions are still satisfied. From this perspective, the main advantage of
the Gaussian system turns out to be that it leads to a normal vector $n^{\mu}$
with coordinate-independent components.


\subsection{Hopf algebras and noncommutative calculus}

We now introduce the basic notion of Hopf algebras and the associated
noncommutative calculus \cite{aschieri1}. We will define only those objects
that will be necessary for our analysis.

\subsubsection{Hopf algebras}

Let us start by introducing the vector space $\mathbb{K}$ of smooth real or
complex vector fields on our classical (commutative) differentiable manifold
$\mathcal{M}$. One can always equip $\mathbb{K}$ with a Lie bracket $[u, v]$
which obeys the Jacobi identity. The pair $\mathcal{A} := (\mathbb{K}, [\cdot
, \cdot])$ is the Lie algebra of classical infinitesimal diffeomorphisms on
$\mathcal{M}$. Infinitesimal transformations of tensors under diffeomorphisms
are provided by the Lie derivative $\mathcal{L}_v$ which obeys $\mathcal{L}_v
\circ \mathcal{L}_u - \mathcal{L}_u \circ \mathcal{L}_v = \mathcal{L}_{[v,u]}$
where $\circ$ stands for composition. 

The Lie derivative of a tensor produces a tensor of the same type and
weight. We shall see in Section~\ref{s:Deformed} that $\star$-diffeomorphisms
obeying the standard Leibniz rule do not satisfy such a condition. We will
therefore be led to a suitable modification of comultiplication. Classically,
infinitesimal diffeomorphisms act on tensor products of tensor fields, $\tau
\otimes \tau'$, by means of the Leibniz formula $\mathcal{L}_v (\tau \otimes
\tau') = (\mathcal{L}_v \tau ) \otimes \tau' + \tau \otimes (\mathcal{L}_v
\tau' ) $. This equation can be interpreted as using the representation
$v\mapsto \mathcal{L}_v$ of vector fields as Lie derivatives after applying
comultiplication $v\mapsto v\otimes 1+1\otimes v$. Moreover, one can define
inverse infinitesimal diffeomorphisms by $v \, \rightarrow \, -v$ and
interpret the complex unit $1\in K^{\otimes 0}$ as a neutral element which
acts by $\mathcal{L}_1 \equiv 1$.

These are the ingredients which can be generalized to a Hopf algebra. To this
end, for an abstract Lie algebra $(\mathbb{K},[\cdot,\cdot])$, one constructs
the universal enveloping algebra $U\mathbb{K}$ (also denoted as
$U[\mathcal{A}]$) as the quotient $\mathcal{F}/\mathcal{I}$, where
$\mathcal{F}$ is the free algebra generated by $(\mathbb{K},\otimes)$ and
$\mathcal{I} \subset\mathbb{F}$ the subspace containing all elements of the
form $u \otimes v - v \otimes u - [u,v]$. The Leibniz rule
is then related to a coalgebra structure.  In the example of infinitesimal
diffeomorphisms, the Leibniz rule gives us the action of $\mathcal{A}$ on
tensor products of functions on $\mathcal{M}$. Abstractly, we can write this
action as the result of a coproduct on $U[\mathcal{A}]$, given by an algebra
homomorphism $\Delta\colon U[\mathcal{A}] \mapsto U[\mathcal{A}]\otimes
U[\mathcal{A}]$.  The universal enveloping algebra of a Lie algebra
$\mathcal{A}$ has a trivial coproduct given by $\Delta v = v \otimes 1 + 1
\otimes v$ for any $v \, \in \, \mathbb{K}$. If $U[\mathcal{A}]$ is instead
equipped with a different coproduct, it is called a Hopf algebra, or quantum
Lie algebra, provided that the following conditions hold: (i) Comultiplication
is coassociative: $(\Delta\otimes 1)\circ\Delta= (1\otimes\Delta)\circ\Delta$.
(ii) There is an inversion map or antipode $S\colon U[\mathcal{A}]\to
U[\mathcal{A}]$ which is an antihomomorphism. (iii) The unit (or neutral)
element $\mathbb{I}\in U[\mathcal{A}]$ is complemented by a co-unit $\epsilon
\colon U[\mathcal{A}] \mapsto \mathbb{C}$ which is a homomorphism. (iv) These
maps are compatible with the multiplication map $\mu \colon
U[\mathcal{A}]\otimes U[\mathcal{A}] \to U[\mathcal{A}]$ in the sense that
$\mu \circ (S \otimes 1 ) \circ \Delta = \mu \circ ( 1 \otimes S) \circ \Delta
= \mathbb{I} \epsilon$. If these conditions are satisfied, the quintuple $H =
(U[\mathcal{A}], \mu, \Delta, \epsilon, S)$ constitutes a Hopf algebra.  For
the universal enveloping algebra of a Lie algebra, for instance, we have $S(v)
= -v$ and $\epsilon(v) = 0$ for $v\in\mathcal{A}$, as well as
$S(\mathbb{I})=\mathbb{I}$ and $\epsilon(\mathbb{I}) = 1$.

It is possible to construct a Hopf algebra from a Lie algebra by using
Drinfeld twists \cite{drin,DT}. The Hopf algebra of 4-dimensional
diffeomorphisms has been studied in Refs.~\cite{aschieri1,aschieri2}. In the
present work we are interested in deriving the deformation theory of the
hypersurface Lie algebroid generating (3+1)-dimensional diffeomorphisms, as
reviewed in the preceeding section for classical differential calculus. To
this end, we derive the Hopf-algebra relations of 4-dimensional
diffeomorphisms explicitly for the specific case of the Moyal-Weyl
noncommutative spacetime (or $\theta$-canonical space).

\subsubsection{Noncommutative calculus}

Suppose that space-time coordinates (locally) obey a Heisenberg-like
commutation relation, such that the commutator between coordinates is
analogous to the commutation relation between configuration and momentum
variables in quantum mechanics:
\begin{equation}\label{canonical}
[\widehat{x}^\mu, \widehat{x}^\nu ] = i \theta^{\mu\nu} \, .
\end{equation}
We restrict our attention to the case in which $\theta^{\mu\nu} =
-\theta^{\nu\mu}$ is constant and real. (It is an anti-symmetric matrix of
numbers and does not depend on coordinate operators.) This is the so-called
Moyal-Weyl spacetime \cite{moy}. As a result of assuming such a non-trivial
commutator, the multiplication between functions no longer enjoys the
commutativity property:
\begin{equation}
F(\widehat{x})G(\widehat{x}) \neq G(\widehat{x})F(\widehat{x})\, .
\end{equation}
By means of a Moyal-Weyl map $\Omega$ \cite{moy}, it is possible to establish
a correspondence between the object $F(\widehat{x})G(\widehat{x})$ and a
suitably modified multiplication rule $f(x) \star g(x)$ between functions of
coordinates,
\begin{equation}
F(\widehat{x})G(\widehat{x}) =: \Omega( f(x) \star g(x)) \, .
\end{equation}
One can show that there are infinitely many possible choices for $\Omega$ that
reproduce standard expressions in the appropriate limit. Thus, given a
noncommutative algebra there is no unique Weyl map.

For the constant-$\theta$ case, the most straightforward choice is
\begin{equation}
\label{moyal}
f(x) \star g(x) = f(x)
e^{-\frac{1}{2}i\overleftarrow{\partial_\alpha}\theta^{\alpha\beta
  }\overrightarrow{\partial_\beta}}g(x)\,.
\end{equation}
We follow the usual quantum-group notation and introduce the twist element
$\mathcal{F} = f^\alpha \otimes f_\alpha :=
e^{\frac{1}{2}i\theta^{\alpha\beta}\partial_\alpha \otimes \partial_\beta} \in
U[{\cal A}]\otimes U[{\cal A}]$ and its inverse, $\mathcal{F}^{-1} =
\overline{f}^\alpha \otimes \overline{f}_\alpha :=
e^{-\frac{1}{2}i\theta^{\alpha\beta}\partial_\alpha
  \otimes \partial_\beta}$. Here, $\alpha$ is used as a multi-index as shown
by an expansion of the exponential function:
\begin{eqnarray}
 {\cal F}&=&1+\frac{1}{2}i\;\theta^{\alpha\beta}\partial_{\alpha}\otimes
\partial_{\beta}- \frac{1}{8}
  \theta^{\alpha_1\beta_1}\theta^{\alpha_2\beta_2} \partial_{\alpha_1}
 \partial_{\alpha_2}\otimes \partial_{\beta_1}\partial_{\beta_2}+\cdots\nonumber\\
&&+
 \frac{1}{n!} (i/2)^n\theta^{\alpha_1\beta_1}\cdots
 \theta^{\alpha_n\beta_n} \partial_{\alpha_1} \cdots
 \partial_{\alpha_n}\otimes \partial_{\beta_1}\cdots\partial_{\beta_n}+\cdots\,.
\end{eqnarray}
We can then write
\begin{equation}
 f_{\alpha} = \sum_{n=0}^{\infty}
 \frac{(i/2)^{n/2}}{\sqrt{n!}} \partial_{\alpha_1}\cdots\partial_{\alpha_n}\,,
\end{equation}
raise the multi-index using
$\theta^{\alpha_1\beta_1}\cdots\theta^{\alpha_n\beta_n}$, and write more
compactly
\begin{equation} \label{fstarg}
f(x) \star g(x) =:  \overline{f}^\alpha(f(x))\overline{f}_\alpha(g(x)) \, .
\end{equation}
Thus, the identity or neutral element of the tensor product of algebras,
$U[\mathcal{A}]\otimes U[\mathcal{A}]$, is given by $1 \otimes 1 =
\mathcal{F}^{-1} \mathcal{F} = \overline{f}^\beta f^\alpha \otimes
\overline{f}_\beta f_\alpha$. In this notation, when we omit the right (or
left) arrow over partial derivatives $\overrightarrow{\partial}_\alpha$ (or
$\overleftarrow{\partial}_\alpha$), the derivative on the left-hand side of a
tensor product acts to the left while the derivative on the right-hand side
acts on functions standing to the right of the star.

The $\star$-product allows one to map the product of operator-valued functions
to a modified product between functions. The product is noncommutative but
still obeys associativity:
\begin{equation}
(f \star g) \star h = f \star (g \star h) \, .
\end{equation}
In terms of the twist and the coproduct, the associative property can be
expressed as
\begin{equation}
\mathcal{F}_{12}(\Delta \otimes 1) \mathcal{F} = \mathcal{F}_{23}(1\otimes
\Delta) \mathcal{F} \, , 
\end{equation}
or equivalently
\begin{equation}
\label{asso}
f^\beta f^\alpha_1 \otimes f_\beta f^\alpha_2 \otimes f_\alpha = f^\alpha
\otimes f_\alpha^1 f^\beta \otimes f_\beta f_\alpha^2 \, . 
\end{equation}
In the former equation we have used $\mathcal{F}_{12} = \mathcal{F} \otimes 1
= f^\alpha \otimes f_\alpha \otimes 1 \in U[{\cal A}]\otimes U[{\cal
  A}]\otimes U[{\cal A}]$ and $\mathcal{F}_{23} = 1\otimes \mathcal{F} =
1\otimes f^\alpha \otimes f_\alpha \in U[{\cal A}]\otimes U[{\cal A}]\otimes
U[{\cal A}]$. An analogous property holds for the inverse twist element.
(These identities can be confirmed by using the explicit expression for the
twist $\mathcal{F} = e^{\frac{i}{2}\theta^{\alpha\beta}\partial_\alpha
  \otimes \partial_\beta}$ and its inverse $\mathcal{F}^{-1} =
e^{-\frac{i}{2}\theta^{\alpha\beta}\partial_\alpha \otimes \partial_\beta}$.)
A second property which $\mathcal{F}$ has to satisfy is
\begin{equation}
(\epsilon \otimes 1) \circ \mathcal{F} = 1 = (1 \otimes \epsilon ) \circ
\mathcal{F} \, . 
\end{equation}

We now wish to define a commutator element in $U[\mathcal{A}]\otimes
U[\mathcal{A}]$, which is called the R-matrix and allows us to make a
permutation of the functions we are (star) multiplying. We define
\begin{equation}
\label{Rmat}
f \star g =: \overline{R}^\alpha (g) \star \overline{R}_\alpha(f) \, ,
\end{equation}
where $R^{-1} = \overline{R}^\alpha \otimes \overline{R}_\alpha$. In order to
find the R-metrix in explicit form, we write
\begin{equation}
\begin{split}
f \star g = \overline{f}^\alpha(f) \overline{f}_\alpha(g) = \overline{f}_\beta
f_\gamma \overline{f}^\alpha(f)  \overline{f}^\beta f^\gamma
\overline{f}_\alpha(g) = \overline{f}^\beta (f^\gamma \overline{f}_\alpha(g))
\overline{f}_\beta(f_\gamma \overline{f}^\alpha(f) ) \\ 
= \overline{f}^\beta ( \overline{R}^\alpha (g))  \overline{f}_\beta(
\overline{R}_\alpha(f)) =  \overline{R}^\alpha (g) \star
\overline{R}_\alpha(f) \, , 
\end{split} 
\end{equation}
with $ \overline{R}^\alpha \otimes \overline{R}_\alpha := f^\gamma
\overline{f}_\alpha \otimes f_\gamma \overline{f}^\alpha$. Here we used only
the representation of the identity in the second step. As a result, the
R-matrix is given by $R = R^\alpha \otimes R_\alpha = f_\gamma
\overline{f}^\alpha \otimes f^\gamma \overline{f}_\alpha$. In particular, for
the Moyal-Weyl spacetime we are considering here, one can verify
\begin{equation}
R =  e^{i\theta^{\alpha\beta}\partial_\alpha \otimes \partial_\beta} \, ,
\quad R^{-1} =  e^{-i\theta^{\alpha\beta}\partial_\alpha
  \otimes \partial_\beta}\, . 
\end{equation}
Using twist properties, the Yang-Baxter equation $R_{12}R_{13}R_{23} =
R_{23}R_{13}R_{12}$ follows.

\subsubsection{Twisted and deformed diffeomorphisms}

Before turning to diffeomorphisms, we introduce the notion of a Lie
bracket. We define two different generalizations of standard brackets between
two fields: the $\star$-Lie bracket $[,]_\star$ and the Moyal bracket
$[\mb]$. In the next sections, we will define the action of twisted and
deformed diffeomorphisms on single fields by using these two brackets.  The
$\star$-Lie bracket between two generic vector fields, $v_1$ and $v_2$, is
defined as
\begin{equation}
\label{starBracket}
[v_1,v_2]_{\star} := v_1\star v_2 -  \overline{R}^\alpha (v_2) \star
\overline{R}_\alpha(v_1) \, . 
\end{equation}
In components,
\begin{equation}
[v_1,v_2]_{\star}^{\mu} = v_1^{\rho}\star \partial_{\rho}v_2^{\mu}-
f^{\gamma}\overline{f}_{\alpha}
v_2^{\rho}\star \partial_{\rho}f_{\gamma}\overline{f}^{\alpha} v_1^{\mu}\,.
\end{equation}

Given this definition we can show that
\begin{equation}
[v_1,v_2]_{\star} = [\overline{f^\alpha}(v_1), \overline{f}_\alpha(v_2)] \, ,
\end{equation}
where on the right-hand side we have the classical Lie bracket: We compute
 \begin{equation}
 \begin{split}
[v_1,v_2]_{\star} = v_1\star v_2 -  \overline{R}^\alpha (v_2) \star
\overline{R}_\alpha(v_1) = \overline{f^\alpha}(v_1)\overline{f}_\alpha(v_2) -
f^\gamma \overline{f}_\alpha \overline{f}^\beta (v_2) f_\gamma
\overline{f}^\alpha \overline{f}_\beta(v_1)\\ =
\overline{f^\alpha}(v_1)\overline{f}_\alpha(v_2) - \overline{f}_\alpha(v_2)
\overline{f}^\alpha(v_1) =  [\overline{f^\alpha}(v_1),
\overline{f}_\alpha(v_2)] \, . 
\end{split}
\end{equation}
This $\star$-Lie bracket satisfies the following modification of the Jacobi
identity
\begin{equation}
[v_1,[v_2,v_3]_\star]_\star = [[v_1,v_2]_\star,v_3]_\star +
[\overline{R}^\alpha(v_2), [\overline{R}_\alpha(v_1),v_3]_\star]_\star \, . 
\end{equation}

Alternatively, we can define what we call the Moyal bracket:
\begin{equation}
\label{moyalB}
[v_1\mb v_2] := v_1\star v_2 - v_2\star v_1 \, .
\end{equation}
It obeys the usual Jacobi identity
\begin{equation}
[v_1 \mb [v_2 \mb v_3]] = [[v_1 \mb v_2] \mb v_3] + [v_2 \mb [v_1 \mb v_3] ] \, ,
\end{equation}
in contrast to $\star$-Lie brackets. Indeed, it is immediate to notice that
$[v_1,v_2]_\star \neq [v_1\mb v_2]$. This result will be at the root of the
difference between twisted diffeomorphisms and deformed diffeomorphisms. We
anticipate that the former do not change the action on single fields but have
a modified Leibniz rule, while the latter retain the Leibniz rule but act on
single fields in a non-standard way. As mentioned, to have a consistent
differential structure, we will then have to change the definition of deformed
diffeomorphisms in such a way that there is a deformation not only of the
action but also of the Leibniz rule. We also mention that the Moyal bracket
allows us to map Eq.~\eqref{canonical} into $[ x^\mu \mb x^\nu ] =
i\theta^{\mu\nu}$. Thus, this bracket is needed to provide a representation of
Eq.~\eqref{canonical} on manifolds equipped with the non-standard product of
Eq.~\eqref{moyal}.

Another property which we will extensively use is $\partial_\mu \star f
= \partial_\mu f$, which is a direct consequence of Eq.~\eqref{moyal} with
constant $\theta$, and, consequently, $\partial_\mu (f\star g) =( \partial_\mu
f)\star g + f\star (\partial_\mu g)$.  Finally, as first discussed for
instance in Ref.~\cite{aschieri1}, the $\star$-tensor product of tensors,
which is needed to have a noncommutative differential calculus together with
the generalizations of Lie brackets defined above, is given by
\begin{equation}
\tau \otimes_{\star} \tau' = \overline{f}^\alpha (\tau) \otimes
\overline{f}_\alpha (\tau') \, .
\end{equation}
The tensor product is therefore twisted just as the pointwise product of
functions.

Let us now discuss the two different paths to treating diffeomorphisms on
$\mathcal{A}$, that is twisted and deformed (or $\star$-) diffeomorphisms. As
already stressed, for the latter, which we here study for the first time, we
will consider two different candidates: either with trivial or non-trivial
co-product. The general idea consists in finding a meaningful generalization
of general covariance to noncommutative manifolds, where noncommutative
manifolds are quantizations of classical smooth manifolds in the sense that
the product of fields evaluated at a spacetime point is noncommutative and is
given by the $\star$-product.


\section{Twisted diffeomorphisms}
\label{s:Twisted}

We return to the derivation of hypersurface-deformation brackets, but
now in a generalization to noncommutative calculus.

\subsection{Lie derivative}

We start by analyzing twisted diffeomorphisms, which have already been
introduced, for instance in Ref.~\cite{aschieri1}, in their 4-dimensional
form. Here, we will focus on their $3+1$-dimensional version. We shall see
that most of the statements made in Ref.~\cite{aschieri1} apply also for the
twisting of hypersurface-deformation brackets. 

Consider a generic tensor $u$. On a commutative space, it transforms as $u' =
u + \delta_{v}u = u + \mathcal{L}_v u$ under infinitesimal diffeomorphisms
generated by the vector field $v = v^{\mu}\partial_\mu$. As usual,
$\mathcal{L}_v u$ is the Lie derivative of $u$ along $v$. It is possible to
represent standard diffeomorphisms on $\mathcal{A}$ by means of
\textit{twisting}. For a function
$u$, we write
\begin{equation}\label{deltav}
\delta_{v}u = \mathcal{L}_v u = v^{\rho} \partial_\rho u = f^\beta
\overline{f}^\alpha (v^{\rho} \partial_\rho) f_\beta \overline{f}_\alpha (u)
= (f^\beta (v^{\rho} \partial_\rho) 
f_\beta) \star u = \mathcal{L}_{v^\star} \triangleright u \,  
\end{equation}
We have inserted the representation of the identity in terms of
the twist and its inverse, and defined
\begin{equation} \label{vstar}
v^{\star} := f^\beta(v)f_\beta = \sum_n
\left(-\frac{i}{2}\right)^n\frac{1}{n!} \theta^{\mu_1\nu_1} \dots
\theta^{\mu_n \nu_n}  (\partial_{\mu_1} \dots \partial_{\mu_n}
v^\rho) \partial_{\nu_1} \dots \partial_{\nu_n} \partial_\rho 
\end{equation}
as an element of $U[{\cal A}]$. The application of ${\cal
  L}_{v^{\star}}$ is what we call an infinitesimal \textit{twisted
  diffeomorphism}.

For a vector field $u^{\mu}$, we proceed in a similar way and write
\begin{eqnarray}
 {\cal L}_vu^{\mu} &=& v^{\rho}\partial_{\rho}u^{\mu}-
 (\partial_{\rho}v^{\mu}) u^{\rho}\nonumber\\
&=& f^{\beta}(v^{\rho}\partial_{\rho})f_{\beta}\star
u^{\mu}- \partial_{\rho}(f^{\beta}(v^\mu)f_{\beta})\star u^{\rho}\nonumber\\
&=& (v^{\rho}\partial_{\rho})^{\star}\star
u^{\mu}-(\partial_{\rho}v^{\star})^{\mu} \star u^{\rho}\,, 
\end{eqnarray}
always keeping $v$ to the left of $u$. In the second term, we may change the
ordering by applying the $R$-matrix,
\begin{eqnarray}
{\cal L}_v u^{\mu} &=& v^{\star}\star u^{\mu}-
\bar{R}^{\alpha}(u^{\rho})\star \partial_{\rho}
\bar{R}_{\alpha}(v^{\star})^{\mu}\\
&=& [v^{\star},u]_{\star}\,,
\end{eqnarray}
in order to derive a relationship with Eq.~\eqref{starBracket}. However, this
notation has to be treated with some care because $(v^\star)^{\mu}$ is not a
function but acts to the left on $u^{\rho}$ in the second term of the
commutator.

The same procedure can be used to derive the Lie derivative of an arbitrary
tensor (density), rewriting the classical relationships in such a way that
components of $v$ (the vector field along which we take the Lie derivative)
always stay on the left. For instance, for the metric tensor $g_{\mu\nu}$, we
have
\begin{equation} \label{Lvg}
 {\cal L}_vg_{\mu\nu} = v^{\star}\star g_{\mu\nu}+
 (\partial_{\mu}v^{\star\rho})\star g_{\rho\nu}+
 (\partial_{\nu}v^{\star\rho})\star g_{\mu\rho}\,.
\end{equation}

\subsection{Twisted Gaussian system}

With these preparations, we can introduce the notion of a noncommutative
Gaussian system for twisted diffeomorphisms. From the point of view of
hypersurface deformations, the main property of a Gaussian system should be
that it leads to constant components $g_{0\mu}$ of the metric. In this way,
the lapse function and shift vector in the background metric are fixed, and it
becomes possible to isolate the role of lapse and shift as generators of
hypersurface deformations. The simplest choice of constant background lapse
and shift that is compatible with a non-degenerate metric of Lorentzian
signature is $g_{00}=-1$ and $g_{0i}=0$ for $i\not=0$.

We need to show that there is a choice of coordinates on a noncommutative
manifold such that the metric is Gaussian in the specified sense. We do so by
assuming the classical Gaussian system under the standard product of functions
or coordinates, and showing that there is a frame in which the required
properties are satisfied also for a noncommutative product and twisted
diffeomorphisms. In particular, the classical system provides us with a time
coordinate $t$ such that $n={\rm d}t$ is the co-normal to spatial
hypersurfaces $t={\rm constant}$. The same 1-form is a co-normal on a
noncommutative manifold with twisted diffeomorphisms: For a vector field $X$
tangential to a spatial hypersurface and $n={\rm d}t$, we have
\begin{equation} \label{Xstarn}
 X^{\mu}\star n_{\mu}=i_{X^\star}\star {\rm d}n = \mathcal{L}_{X^\star}
 \triangleright t = X^{\mu}\partial_{\mu}t=0 \,.
\end{equation}
The Lie derivative along $X^{\star}$ is equal to the classical Lie derivative
because all higher-derivative terms in (\ref{vstar}) vanish when acting on a
linear function such as $t$. In a Gaussian frame, the co-normal therefore has
constant components, and so does the normal $n^{\mu}=g^{\mu\nu}\star
n_{\mu}=g^{\mu\nu}n_{\mu}$ because higher derivatives in the star product
vanish when applied to a constant $n_{\mu}$. Here we introduced the inverse
 metric $g^{\nu\alpha}\star g_{\alpha\mu} = \delta^{\alpha}_{\mu}$, defining the
 inverse metric by its action from the left (alternatively one can define 
 the inverse by-the-right) \cite{aschieri1}. 

The normal is therefore normalized with respect to the noncommutative system,
in the following sense:
\begin{eqnarray} \label{Normalization}
i_{n^\star} \star g\star i_n &=& n^{\star\mu}\star g_{\mu\nu}\star n^{\nu}=
 f^{\alpha}n^{\mu}f_{\alpha} \star g_{\mu\nu}\star n^{\nu}\\
 &=&  n^{\mu}g_{\mu\nu}\star n^{\nu}= n_{\nu}\star n^{\nu} = n_{\nu}n^{\nu}=-1\,.
\end{eqnarray}
In a classical Gaussian system, we have $n^{\mu}\nabla_{\mu}n^{\nu}=0$ because
worldlines normal to spatial hypersurfaces are geodesics. In a Gaussian frame,
all contributions from connection components in this equation are zero because
the only relevant ones, 
\begin{equation}
 \Gamma^0_{0\mu}= \frac{1}{2}g^{0\alpha}(\partial_\mu
 g_{0\alpha}+ \partial_0 g_{\mu\alpha}- \partial_\alpha g_{0\mu})=0\,,
\end{equation}
vanish identically for a Gaussian metric. The equation
$n^{\mu}\nabla_{\mu}n^{\nu}=0$ is therefore equivalent to
$n^{\mu}\partial_{\mu}n^{\nu}=0$ in a Gaussian system. The same equation is
true in the form $n^{\mu}\star \partial_{\mu}n^{\nu}=0$ for a noncommutative
Gaussian system because, as we just showed, the components of $n^{\mu}$ are
still constant. From this equation, we can derive $n^{\mu}\star
\nabla_{\mu}\star n^{\nu}=0$ using the definition of the noncommutative
Christoffel connection from \cite{aschieri1}, which gives
\begin{eqnarray}
 \Gamma^0_{0\mu} &=& \frac{1}{2}g^{0\alpha}\star (\partial_\mu
 g_{0\alpha}+ \partial_0 g_{\mu\alpha}-\partial_\alpha g_{0\mu})\\
 & =& \frac{1}{2} g^{0\alpha}\star \partial_0 g_{\mu\alpha} = 0 \,, 
\end{eqnarray}
for the relevant connection components. 

It will be convenient to do calculations of the hypersurface-deformation
brackets in a Gaussian frame. However, whenever possible, we will not make
explicit use of the fact that normal components are constant in order to
display all relevant star products. In particular, in order to be as general
as possible, we will derive differential equations for the normal and
tangential components of a Gaussian vector field without using constant
components of the normal. We will see that a counterterm is then required in
the classical Gaussian condition. We then analyze these differential equations
using all the properties of a Gaussian frame, including the constant nature of
components of the normal. This step will allow us to show that there is a
well-posed initial-value problem and a set of algebroid brackets which depend
only on hypersurface data.

\subsection{Gaussian condition}

We are interested in deriving properties of hypersurface deformations in
noncommutative space-time, with possible modifications of the action of
twisted diffeomorphisms. We modify the classical expression used to define a
Gaussian vector field as follows: Instead of $i_n \mathcal{L}_v g = 0$, we
require that
\begin{equation}
\label{twistGauss}
 \left(\mathcal{L}_{v^\star}   \triangleright g\right)\star i_n = 0 \, . 
\end{equation}
We act with $i_n$ from the right in order to make sure that it stands next to
the metric, without components of $v^{\star}$ in between.  Classically, we say
that $v$ is Gaussian if a diffeomorphism of the metric along the direction
given by $v$ does not have a normal component. We have generalized this
statement by saying that the twisted infinitesimal diffeomorphism of $g$,
generated by $v$, gives zero if we $\star$-contract the result with the normal
$n$. Since the normal components are constant, (\ref{twistGauss}) is
equivalent to the classical condition on Gaussian vector fields, and it is
therefore consistent with the metric form of a Gaussian system.

We have that $ i_n \mathcal{L}_v g = n^\mu ( \mathcal{L}_v g)_{\mu\nu}$, and
analogously we can write the twisted version in components as
$\left(\mathcal{L}_{v^\star} \triangleright g\right)_{\mu\nu} \star n^\mu$,
where the Lie derivative of the metric is given in (\ref{Lvg}) in terms of
twisted diffeomorphisms. We rewrite star products using (\ref{fstarg}), for
instance $(v^\rho)^\star \star \partial_\rho g=
\overline{f}^\alpha((v^\rho)^\star \partial_\rho) \overline{f}_\alpha(g)$ in
the first term, and therefore obtain the Gaussian condition for $v$ as
\begin{equation}
\label{defGauss}
\left( \mathcal{L}_{\overline{f}^\alpha(v^\star) \overline{f}_\alpha} g
\right) \star i_n = 0 \, . 
\end{equation}
The next step is to try and obtain relations for the normal and tangential
components of the $\star$-Lie bracket between the normal $n$ and the Gaussian
vector field $v$. In doing that, we will try to follow as close as possible
the steps of the derivation for the commutative case.

First, we would like to compute
$\mathcal{L}_{\overline{f}^\alpha(v^{\star}) \overline{f}_\alpha}( g \star
i_n)$, or the action of the twisted Lie derivative on the $\star$-product of
two fields:
\begin{equation}
\begin{split}
 \overline{f}^\alpha(v^{\star}) \overline{f}_\alpha( g \star i_n) =
\overline{f}^\alpha(v^{\star}) \overline{f}_\alpha(\overline{f}^\beta(
g)\overline{f}_\beta( i_n)) \\ =
\overline{f}^\alpha(v^{\star})\overline{f}^1_\alpha \overline{f}^\beta (g)
\overline{f}^2_\alpha \overline{f}_\beta(i_n) =
\overline{f}^\alpha(v^\mu)^{\star}\overline{f}^1_\alpha \overline{f}^\beta
(\partial_\mu g_{\sigma \nu}) \overline{f}^2_\alpha \overline{f}_\beta(n^\sigma) \\+
\overline{f}^\alpha(\partial_\nu v^\mu)^{\star}\overline{f}^1_\alpha
\overline{f}^\beta 
( g_{\sigma \mu}) \overline{f}^2_\alpha \overline{f}_\beta(n^\sigma)+
\overline{f}^\alpha(v^\mu)^{\star}\overline{f}^1_\alpha \overline{f}^\beta (g)
\overline{f}^2_\alpha \overline{f}_\beta( i_{\partial_\mu n})  \, .
\end{split}
\end{equation}
Adding and subtracting the term $ \overline{f}^\alpha(\partial_\sigma
v^\mu)^{\star}\overline{f}^1_\alpha \overline{f}^\beta ( g_{\nu \mu})
\overline{f}^2_\alpha \overline{f}_\beta(n^\sigma)$, we obtain
\begin{equation} \label{terms}
\begin{split}
\overline{f}^\alpha(v^\mu)^{\star}\overline{f}^1_\alpha \overline{f}^\beta
(\partial_\mu g_{\sigma \nu}) \overline{f}^2_\alpha \overline{f}_\beta(n^\sigma) \\+
\overline{f}^\alpha(\partial_\nu v^\mu)^{\star}\overline{f}^1_\alpha
\overline{f}^\beta 
( g_{\sigma \mu}) \overline{f}^2_\alpha
\overline{f}_\beta(n^\sigma)+\overline{f}^\alpha(\partial_\sigma
v^\mu)^{\star}\overline{f}^1_\alpha \overline{f}^\beta 
( g_{\nu \mu}) \overline{f}^2_\alpha
\overline{f}_\beta(n^\sigma)\\-\overline{f}^\alpha(\partial_\sigma
v^\mu)^{\star}\overline{f}^1_\alpha \overline{f}^\beta 
( g_{\nu \mu}) \overline{f}^2_\alpha \overline{f}_\beta(n^\sigma)+
\overline{f}^\alpha(v^\mu)^{\star}\overline{f}^1_\alpha \overline{f}^\beta (g)
\overline{f}^2_\alpha \overline{f}_\beta( i_{\partial_\mu n})  \, .
\end{split}
\end{equation}
Using both \eqref{asso} and \eqref{Rmat}, for the first three terms we have 
\begin{equation}
\begin{split}
 \overline{f}^\alpha(v^\mu)^{\star}\overline{f}^1_\alpha \overline{f}^\beta
(\partial_\mu g_{\sigma \nu}) \overline{f}^2_\alpha \overline{f}_\beta(n^\sigma) +
\overline{f}^\alpha(\partial_\nu v^\mu)^{\star}\overline{f}^1_\alpha
\overline{f}^\beta 
( g_{\sigma \mu}) \overline{f}^2_\alpha \overline{f}_\beta(n^\sigma)\\+\overline{f}^\alpha(\partial_\sigma v^\mu)^{\star}\overline{f}^1_\alpha \overline{f}^\beta
( g_{\nu \mu}) \overline{f}^2_\alpha \overline{f}_\beta(n^\sigma)=
 \overline{f}^\alpha_1 \overline{f}^\beta(v^\mu)^{\star}\overline{f}^\alpha_2
 \overline{ f}_\beta (\partial_\mu g) \overline{f}_\alpha (i_n)\\+
\overline{f}^\alpha_1  \overline{f}^\beta(\partial_\nu v^\mu)^{\star}\overline{f}^\alpha_2
\overline{f}_\beta( g_{\sigma \mu}) \overline{f}_\alpha (n^\sigma)+\overline{f}^\beta \overline{f}^\alpha_1(\partial_\sigma v^\mu)^{\star}\overline{f}^\alpha_2 \overline{f}_\beta
( g_{\nu \mu}) \overline{f}_\alpha (n^\sigma)\\
= \left(
   \mathcal{L}_{\overline{f}^\alpha(v^{\star})\overline{f}_\alpha}g\right)\star
 i_n\,.
\end{split}
\end{equation}
We write the last two terms of (\ref{terms}) as
\begin{equation}
\begin{split}
 \overline{f}^\alpha(v^\mu)^{\star}\overline{f}^1_\alpha \overline{f}^\beta (g)
\overline{f}^2_\alpha \overline{f}_\beta(i_{\partial_\mu n})-\overline{f}^\alpha(\partial_\sigma v^\mu)^{\star}\overline{f}^1_\alpha \overline{f}^\beta
( g_{\nu \mu}) \overline{f}^2_\alpha \overline{f}_\beta(n^\sigma)\\
=
\overline{f}^\alpha ( \overline{R}^\gamma(g)) \overline{f}^1_\alpha
 \overline{f}^\beta
 \overline{R}_\gamma(v^\mu)^{\star})\overline{f}_\beta
\overline{f}^2_\alpha( i_{\partial_\mu n})-\overline{f}^\alpha(\overline{R}^\gamma(g_{\nu\mu}))\overline{f}^1_\alpha \overline{f}^\beta
\overline{R}_\gamma(\partial_\sigma v^\mu)^{\star} \overline{f}^2_\alpha \overline{f}_\beta(n^\sigma) \\ = \overline{f}^\alpha ( \overline{R}^\gamma(g)) \overline{f}_\alpha
 \left( \overline{R}_\gamma((v^\mu)^{\star}\partial_\mu) \star i_n \right)-\overline{f}^\alpha(\overline{R}^\gamma(g_{\nu\mu})) 
\overline{f}_\alpha(\overline{R}_\gamma(\partial_\sigma v^\mu)^{\star} \star (n^\sigma)) \\
 =  \overline{R}^\gamma(g) \star
 \left(i_{\mathcal{L}_{\overline{R}_\gamma(\overline{f}^\beta(v^{\star})\overline{f}_\beta)}n}
 \right) \, , 
\end{split}
\end{equation}
and arrive at
\begin{equation}
\label{qLeibn}
\mathcal{L}_{\overline{f}^\alpha(v^{\star}) \overline{f}_\alpha}( g \star i_n)
=  \left(
  \mathcal{L}_{\overline{f}^\alpha(v^{\star})\overline{f}_\alpha}g\right)\star
i_n + \overline{R}^\alpha(g) \star
\left(i_{\mathcal{L}_{\overline{R}_\alpha(\overline{f}^\beta(v^{\star})\overline{f}_\beta)}n}
\right) \, . 
\end{equation}
We see that, as a direct consequence of loss of commutativity of the
$\star$-product, the Leibniz rule does not apply. It is modified through
the action of the R-matrix, as one could have anticipated. Using the above
expressions we can rewrite Eq. \eqref{defGauss} as
\begin{equation}
\left( \mathcal{L}_{\overline{f}^\alpha(v^{\star}) \overline{f}_\alpha} g
\right) \star i_n = \mathcal{L}_{\overline{f}^\alpha(v^{\star})
  \overline{f}_\alpha}( g \star i_n) - \overline{R}^\alpha(g) \star
\left(i_{\mathcal{L}_{\overline{R}_\alpha(\overline{f}^\beta(v^{\star})\overline{f}_\beta)}n}
\right) = 0 \, . 
\end{equation}

The next step is an application of the Cartan identity. The validity of such
an identity is usually required as an axiom, or assumed (see for
instance \cite{schupp}), but it is possible to prove it in the following
manner. Let us make indices explicit in
\begin{equation}
\begin{split}
\mathcal{L}_{\overline{f}^\alpha(v^{\star}) \overline{f}_\alpha}( g \star i_n)
= \overline{f}^\alpha(v^\rho\partial_\rho)^{\star}
\overline{f}_\alpha\left(g_{\mu\nu}\star n^\mu \right) = (v^\rho)^{\star}
\star \partial_\rho ( g_{\mu\nu} \star n^\mu)  + \partial_\nu (v^\rho)^{\star}
\star (g_{\rho\mu} \star n^\mu) \\ 
= (v^\rho)^{\star} \star \partial_\rho ( g_{\mu\nu} \star n^\mu)
+ \partial_\nu (v^\rho)^{\star} \star (g_{\rho\mu} \star n^\mu) +
(v^\rho)^{\star} \star \partial_\nu ( g_{\rho\mu} \star n^\mu) - (
v^\rho)^{\star} \star \partial_\nu ( g_{\rho\mu} \star n^\mu) \\
= \partial_\nu ( (v^\rho)^{\star} \star g_{\rho\mu} \star n^\mu) +
(v^\rho)^{\star} \star ({\rm d}n)_{\rho \nu} \,  , 
\end{split}
\end{equation}
where we defined the two-form $({\rm d}n)_{\rho \nu } := \partial_\rho ( g_{\mu\nu}
\star n^\mu) - \partial_\nu ( g_{\mu\rho}\star n^\mu)$. Thus, we derived
\begin{equation}
\mathcal{L}_v^{\star} \triangleright (g \star i_n) = i_{v^{\star}} \star
{\rm d}(g\star i_n) + {\rm d}(i_{v^{\star}}\star g\star i_n) \, , 
\end{equation}
commonly known as the Cartan identity.

With this result, we have
\begin{equation}
\begin{split}
 \mathcal{L}_{\overline{f}^\alpha(v^{\star}) \overline{f}_\alpha}( g \star
 i_n) - \overline{R}^\alpha(g) \star
 \left(i_{\mathcal{L}_{\overline{R}_\alpha(\overline{f}^\beta(v^{\star})\overline{f}_\beta)}n}
 \right) \\ =  i_{v^{\star}} \star {\rm d}(g\star i_n) + {\rm
   d}(i_{v^{\star}}\star g\star
 i_n) - \overline{R}^\alpha(g) \star
 \left(i_{\mathcal{L}_{\overline{R}_\alpha(\overline{f}^\beta(v^{\star})\overline{f}_\beta)}n}
 \right) = 0 \, . 
 \end{split}
\end{equation}
Now we use ${\rm d}n ={\rm d}(g\star i_n)= 0$ and obtain
\begin{equation}
\overline{R}^\alpha(g) \star
\left(i_{\mathcal{L}_{\overline{R}_\alpha(\overline{f}^\beta(v^{\star})\overline{f}_\beta)}n}
\right) = {\rm d}(i_{v^{\star}}\star g\star i_n) \, . 
\end{equation}

\subsection{Decomposition}

Decomposing $v^{\star}$ into components normal and tangential to
hypersurfaces, $v^{\star} = (N^{\star}\star n)^{\star} + (M^{\star}\star
X)^{\star}$ (with $N^{\star} := f^{\alpha}(N)f_{\alpha}$ and $M^{\star} :=
f^{\alpha}(M)f_{\alpha}$), we write
\begin{equation}
\overline{R}^\alpha(g) \star
\left(i_{\mathcal{L}_{\overline{R}_\alpha(\overline{f}^\beta(N^{\star}\star
    n)^{\star}\overline{f}_\beta)}n} \right) + \overline{R}^\alpha(g) \star
\left(i_{\mathcal{L}_{\overline{R}_\alpha(\overline{f}^\beta(M^{\star}\star
    X)^{\star}\overline{f}_\beta)}n} \right) = -{\rm d}N^\star \, , 
\end{equation}
where we have used the relations
\begin{equation}
i_{n^\star} \star g\star i_n = -1 \, \quad \, i_{X^\star}\star g\star i_n = 0 \,  ;
\end{equation}
see (\ref{Normalization}).

Writing indices explicitly,
\begin{eqnarray}
\overline{R}^\alpha(g_{\nu\mu}) \star \left[
  \overline{R}_\alpha\overline{f}^\beta(N^\star\star
  n^\rho)^\star\overline{f}_\beta(\partial_\rho n^\mu)-
  \overline{R}_\alpha\overline{f}^\beta\partial_\rho (N^\star\star
  n^\mu)^\star \overline{f}_\beta(n^\rho)\right. \nonumber  \\ 
\left. + \overline{R}_\alpha\overline{f}^\beta(M^\star\star
  X^\rho)^\star\overline{f}_\beta(\partial_\rho n^\mu)-
  \overline{R}_\alpha\overline{f}^\beta\partial_\rho (M^\star\star
  X^\mu)^\star \overline{f}_\beta(n^\rho)  \right] = -\partial_\nu N^\star \,
, 
\end{eqnarray}
where we left implicit only the internal index $a$ in $M\star X^\mu \equiv
M^a \star X^\mu_a$.  Using Eq.~\eqref{Rmat}, we have
\begin{eqnarray}
\overline{R}^\alpha(g_{\nu\mu}) \star \left[
  \overline{R}_\alpha\overline{f}^\beta(N^\star\star
  n^\rho \partial_\rho)^\star\overline{f}_\beta( n^\mu)-
  \overline{f}^\beta(\overline{R}^\gamma(n^\rho \partial_\rho))
  \overline{f}_\beta(\overline{R}_\alpha\overline{R}_\gamma( N^\star\star
  n^\mu)^\star)  \right. \nonumber \\ 
\left.+ \overline{R}_\alpha\overline{f}^\beta(M^\star\star
  X^\rho \partial_\rho)^\star\overline{f}_\beta( n^\mu)-\overline{f}^\beta(
  \overline{R}_\alpha\overline{R}^\gamma( n^\rho\partial_\rho) )
  \overline{f}_\beta(\overline{R}_\gamma(M^\star\star X^\mu)^\star)  \right] =
-\partial_\nu N^\star \, , 
\end{eqnarray}
and finally, recalling Eq.~\eqref{starBracket},
\begin{equation}\label{twnt}
\overline{R}^\alpha(g_{\nu\mu}) \star \left( [\overline{R}_\alpha(N^\star\star
  n)^\star, n]^\mu_\star + [\overline{R}_\alpha(M^\star\star X)^\star,
  n]^\mu_\star \right) = -\partial_\nu N^\star \, . 
\end{equation}

So far, following Refs.~\cite{aschieri1,aschieri2}, we have defined twisted
(four) diffeomorphisms by a representation of the infinitesimal
diffeomorphisms of classical differential manifolds on noncommutative
manifolds or, rather, on manifolds equipped with a non-trivial
$\star$-multiplication rule \eqref{moyal}. As a consequence, they have an
undeformed action on single fields or tensors but, due to the Moyal
$\star$-product, act non-trivially on products of two or more objects. Thus,
twisting diffeomorphisms corresponds to mapping them to the Moyal space (or,
more generally, to a manifold with noncommutative products).  In order to find
formulae relating the lapse function and shift vector components, it will be
more useful to rewrite the relation \eqref{twnt} as one on the commutative
classical manifold in an intermediate step. We will then represent the final
hypersurface-deformation brackets on the Moyal space in order to obtain a
twisted version of the HDA.

Using the definition of the R-matrix as well as that of the $\star$-Lie
bracket, we rewrite Eq.~\eqref{twnt} as
\begin{eqnarray}
  -\partial_\nu N^\star  &=& (Nn^\rho  g_{\nu\mu})\star \partial_\rho n^\mu  -
  (\partial_\rho (Nn^\mu)g_{\nu\mu})\star n^\rho + (M^\rho
  g_{\nu\mu})\star \partial_\rho n^\mu - (\partial_\rho M^\mu g_{\nu\mu})\star
  n^\rho \nonumber\\
  &=& g^\star_{\mu\nu}\star N^\star \star (n^\rho
  \star \partial_\rho n^\mu-(\partial_{\rho}n^{\mu})\star n^{\rho}) -
  g^\star_{\mu\nu} \star n^\mu\star \partial_\rho 
  N^\star \star   n^\rho\label{partialNstar}\\
 && + g^\star_{\nu\mu}\star M^\rho \star\partial_\rho n^\mu
  -  g^\star_{\nu\mu}\star\partial_\rho M^\mu \star n^\rho \,.  \nonumber
\end{eqnarray}
We can now use the constant nature of $n_{\mu}$ in a Gaussian frame, so that
$n^{\rho}$ star-commutes with any function and the partial
gradient $\partial_\rho n^\mu=0$ vanishes. Multiplying both sides of
(\ref{partialNstar}) by $n^\nu$, we have
\begin{equation} \label{ndN}
-n^\nu \star \partial_\nu N^\star - \partial_\nu N^\star \star n^\nu = -n^\nu \star
g^\star_{\nu\mu}\star \partial_\rho M^\mu \star n^\rho  \, , 
\end{equation}
where we also used $n^\mu \star n_\mu = -1$. Applying the product rule in
\begin{equation}
 0=n^{\rho}\star \partial_{\rho}(n_{\mu}\star M^{\mu})=
 (n^{\rho}\star \partial_{\rho}n_{\mu}) \star M^{\mu}+ n_{\mu}\star
 (n^{\rho}\star \partial_{\rho} M^{\mu})+ (n_{\mu}\star n^{\rho}-n^{\rho}\star
 n_{\mu}) \star \partial_{\rho}M^{\mu}\,,
\end{equation}
and using $n^\nu \star X_\nu = 0$ as well as the vanishing star commutator
$n_{\mu}\star n^{\rho}-n^{\rho}\star n_{\mu}=0$ of the constant $n_{\mu}$, implies
that $n^{\rho}\star \partial_{\rho} M^{\mu}=0$.
Thus, we finally
obtain
\begin{equation}
0=-n^\nu \star \partial_\nu N^\star - \partial_\nu N^\star \star
n^\nu=-2n^{\nu}\partial_{\nu}N^{\star} = -2n^\nu \partial_\nu N \,.
\end{equation}
In the last step, we have mapped the expression back to the commutative space
and, therefore, multiplication is the usual commutative rule.

The tangential projection of Eq.~\eqref{twnt} is made in a similar way.  By
$\star$-multiplying with $q^{ab}$, we have
\begin{equation}
[n , M ]^a_\star=q^{ab}\star \partial_b N^\star    \, .
\end{equation}
Lapse $N$ and shift $M^a$ are subject to the same type of partial differential
equations as in the classical derivation. Therefore, they are extendable to a
pace-time neighborhood of a spatial hypersurface and can be used in the Lie
brackets of Gaussian space-time vector fields.

\subsection{Brackets}

We are now ready to evaluate the $\star$-Lie bracket of space-time vector
fields. We calculate the $\star$-product between the $\star$-Lie bracket
$[v^{\star}_{1} , v^{\star}_2]^\mu_\star$ and an arbitrary scalar function $f$
for twisted diffeomorphisms,
\begin{eqnarray} \label{vvGen}
[v^{\star}_{1} , v^{\star}_2]^\mu_\star \star f &=&\left(
  (v^\rho_1)^\star \star \partial_\rho (v^\mu_2)^\star -
  \overline{R}^{\alpha}(v^\rho_2)^\star \star
  \overline{R}_\alpha(\partial_\rho v^\mu_1)^\star \right)
\star \partial_{\mu} f \nonumber\\ 
&=& v^\rho_1 \partial_\rho v^\mu_2 \partial_{\mu}f - \partial_\rho v^\mu_1
v^\rho_2 \partial_{\mu} f\nonumber\\ 
&=&
(N_1 n^\rho + M^\rho_1) \partial_\rho  (N_2 n^\mu + M^\mu_2) \partial_{\mu}f - 
\partial_\rho
(N_1 n^\mu + M^\mu_1)  (N_2 n^\rho + M^\rho_2)  \partial_{\mu}f \nonumber\\ 
&=& (N_1
n^\rho \partial_\rho N_2 - \partial_\rho N_1 N_2 n^\rho ) n^\mu \partial_{\mu}f + (
\mathcal{L}_{M_1 }  N_2 - \mathcal{L}_{M_2}N_1 )  n^\mu \partial_{\mu}f\nonumber\\ 
& &\,\,\,\,+ [M_1 , M_2 ]^\mu \partial_{\mu}f + N_1
[n,M_2]^\mu \partial_{\mu}f- N_2 [n , M_1]^\mu \partial_{\mu} f \, ,
\end{eqnarray}
and extract normal and tangential terms and using the above relations for
$[n,M]$:
\begin{eqnarray}
&&[(0,M^a_1), (0,M^b_2)] = (0, \mathcal{L}_{M_1}M_2) \, , \\
&&[(N,0) , (0,M^a)]  = \left( -\mathcal{L}_{M} N ,0 \right) \, , \\
&&[(N_1,0) , (N_2,0)] = ( 0,   (N_1  \partial_b N_2 -N_2  \partial_b
N_1)q^{ab} ) \, . 
\end{eqnarray}
The fact that this result coincides with Eqs.~\eqref{classHDA} confirms the
claim \cite{aschieri1} that twisted noncommutative gravity with the Moyal
product has the same symmetry algebra as classical GR. Thus, the only
deformations of symmetries are encoded in the coalgebraic sector where, due to
the non-standard multiplication, the Leibniz rule does not apply. Having
a closed and consistent set of brackets also ensures that noncommutative
gravity possesses the same number of degrees of freedom as GR, as one should
expect. We shall see that this statement remains true also for deformed
diffeomorphism symmetries, in which case the HDA does receive $\star$-product
deformations.

Once one has obtained the Poisson brackets for general coordinate
transformations, it is of interest to study their Minkowski (or flat)
limit. In this way, one restricts the set of diffeomorphisms and only allows a
subset of coordinate transformations, which are the isometries of Minkowski
spacetime. In terms of hypersurface deformations, this restriction can be
implemented by using the Euclidean spatial metric and requiring lapse and
shift to be linear in space coordinates, of the form $N = \alpha + \alpha_i
x^i$ and $M^a = \beta^a + R^a_b x^b$ ($R_{ab}$ being a matrix of rotations in
space). The interested reader can take a look at
Refs. \cite{teit,nclqg,phenolqg1} for the Minkowski limit of the HDA and its
deformations. Here, as expected, we find that the twisted HDA has no
deformations compared with the standard version of GR. It is then not
difficult to show that, after the specified restrictions, the resulting
Poincar\'e algebra is also unmodified. On the other hand, one can expect that
the action of Poincar\'e generators on products of functions will be
non-trivial as a result of the presence of a noncommutative multiplication
rule. This is consistent with the known fact that the symmetry algebra dual to
the Moyal-Weyl space-time is the so-called $\theta$-Poincar\'e algebra with
standard commutators but deformed coproducts \cite{tetaP}.


\section{Deformed diffeomorphisms}
\label{s:Deformed}

We first perform the Gaussian analysis for the derivation of brackets by
defining $\star$-diffeomorphisms (or, equivalently, deformed diffeomorphisms)
with a deformed action on single tensors but still respecting the Leibniz
rule. This is done in an attempt to reproduce what has been studied for
noncommutative quantum field theories
\cite{szabo1,douglas1,szabo2,douglas2,ncqft1,ncqft2,ncqft3}, where the
relevant $\star$-action is invariant under $\star$-${\rm U}(1)$ symmetries
obeying the Leibniz rule. Some rather encouraging results are achieved but we
will explain later on why there is a strong reason for abandoning the Leibniz
rule and then working with deformed diffeomorphisms with deformed
comultiplication.

\subsection{Deformed diffeomorphisms with trivial coalgebra}

We define a deformed diffeomorphism by its infinitesimal action
\begin{equation}
\label{defDiff}
\mathcal{L}_v \triangleright u := v^\rho \star \partial_\rho \star u = v^\rho
\star \partial_\rho u \, , 
\end{equation}
on functions.  In the last step we used the fact that, for the
constant-$\theta$ case, the action of the derivative is not modified, that is
$\partial_\mu \star f \equiv \partial_\mu f$.  Deformed diffeomorphisms are
different from twisted ones because $v^\rho \star \partial_\rho u \neq
\delta_v u$ defined in (\ref{deltav}).

A deformed Gaussian system can be defined analogously to a twisted one. The
first place where we used the Lie derivative in the construction of a twisted
Gaussian system was in Eq.~(\ref{Xstarn}). Because it acts on a linear
coordinate function $t$, it remains true if we use the Lie derivative
(\ref{defDiff}) corresponding to deformed rather than twisted
diffeomorphisms. The second place, the introduction of a condition on Gaussian
vector fields, will be discussed soon. But first, we have to insert a warning
about a violation of the standard Leibniz rule for the Lie derivative of
deformed diffeomorphisms as defined so far.

In some sense, one could consider deformed diffeomorphisms the most natural
definition of diffeomorphisms on $\mathcal{A}$. According to
Eq.~\eqref{moyal}, we can obtain diffeomorphisms on $\mathcal{A}$ thanks to
the mapping given by the $\star$-product. Using a Weyl map, we have
\begin{equation}
\widehat{\delta}_V F(\widehat{x}) = V(\widehat{x}) \triangleright
F(\widehat{x}) \mapsto v(x) \star f(x) = v^\rho(x) \star \partial_\rho  f(x)
\, , 
\end{equation}
where the last expression gives us exactly the definition we proposed for
deformed diffeomorphisms, \eqref{defDiff}.  

However, an extension to vector
fields and tensors is non-trivial if we want to preserve the Leibniz rule. For
instance, if we attempt such an extension by postulating that the $\star$-Lie
derivative should agree with the
Moyal bracket \eqref{moyalB},
\begin{equation}
\mathcal{L}_{v_1}\triangleright v_2 = [v_1\mb v_2] \, ,
\end{equation}
for two vector fields $v_1$ and $v_2$, the Leibniz rule is in danger when we
apply the derivative to the product of a function $u$ and a vector field $w$:
\begin{eqnarray}
 \mathcal{L}_{v}\triangleright (u\star w) &=&
 v^{\rho}\star\partial_{\rho}(u\star w^{\mu})- u\star
 w^{\rho}\star\partial_{\rho} v^{\mu}\\
&=& (\mathcal{L}_{v}\triangleright u)\star w^{\mu} + u\star
(\mathcal{L}_{v}\triangleright w) +(v^{\rho}\star u-u\star
v^{\rho})\star \partial_{\rho}w^{\mu}\,.
\end{eqnarray}
The last $\star$-commutator violates the Leibniz rule, but it vanishes when
$u$ is a constant, such as a normal component in our deformed Gaussian
system. We may therefore postpone a detailed discussion of the Leibniz rule
and first return to hypersurface deformations.

At this point, we have the necessary ingredients to develop the Gaussian
analysis for deformed diffeomorphisms, with our general aim of deriving the
hypersurface-deformation brackets they imply.  Due to the $\star$-modification
of the action of these symmetries, it is then natural to expect modifications
of the HDA and, thus, a deformed or $\star$-modification of general
covariance.


\subsection{Modified Gaussian condition}

Recall that we are not interested in the Gaussian system in its own
right, but rather have to make sure that the gauge choice leads to brackets of
space-time vector fields which depend only on hypersurface data. The latter
can then be reinterpreted as Lie-algebroid brackets.  The original Gaussian
condition for the metric reads
\begin{equation}
n^\mu \star \mathcal{L}_v\star g_{\mu\nu} = 0 \, .
\end{equation}
However, it does not lead to a well-defined Lie-algebroid
structure for deformed diffeomorphisms.  We modify it by subtracting
a term which will lead to consistent relations:
\begin{equation}\label{gausC3}
n^\mu \star \mathcal{L}_v\star g_{\mu\nu} - \partial_\gamma (v^\rho \star
n^\mu \star g_{\rho\mu})\star g^{\gamma\alpha}\star n^\beta \star
g_{\alpha\beta} \star n_\nu = 0 \, , 
\end{equation}
is the new $\star$-modified Gaussian condition.  In abstract notation, the
commutative analog of the new condition reads
\begin{equation}
i_n \mathcal{L}_v g = (i_n {\rm d}(i_v i_n g)) n \, , 
\end{equation}
or, in components, 
\begin{equation}
\label{defgf}
n^\mu \mathcal{L}_v  g_{\mu\nu} = n^\rho
 \partial_\rho (g_{\delta\gamma}  n^\gamma  v^\delta)  n_\nu \, . 
\end{equation}
The difference with respect to the usual Gaussian condition is that the
variation of the metric $g$ under a diffeomorphism along the direction
identified by $v$ is non-zero. We are therefore choosing a different gauge
where, instead of being zero, the normal contribution to $ \mathcal{L}_v g$ is
fixed to another specific value. Since the structure of hypersurface
deformations should be gauge independent, we expect the new condition
(\ref{defgf}) to imply the same hypersurface-deformation brackets as derived
in \cite{wein} when applied to the ordinary product. In App.~\ref{a:ModGauss},
we confirm that this modification indeed does not change the result of the
classical calculation for commutative theories. As a brief argument, we can
see that the classical condition can be modified by our counterterm because
the latter is zero when the conditions for lapse and shift that follow from
the original condition are satisfied, in particular when
$0=n^{\rho}\partial_{\rho}(N^2)=i_n{\rm d}(i_vi_n g)$. (The counterterm
vanishes ``on shell.'')

Using the Cartan identity, we write the modified Gaussian condition as
\begin{equation}
  i_{v}\star {\rm d}n + {\rm d}(i_v \star i_n \star g)
  +  i_{[n \mb v]}\star g+ ({\rm d} (i_v\star i_n\star g)\star \overleftarrow{
    i_n})\star n = 0\, ,  
\end{equation}
where $\overleftarrow{i_n}$ highlights the fact that the normal vector is
$\star$-contracted with the tensor on the left of the product, ${\rm d}
(i_v\star i_n\star g)$. Decomposing $v=N\star n+M\star X$ and using ${\rm
  d}n=0$ as well as the orthogonality conditions
\begin{equation}\label{ort2}
i_n \star g \star i_n = -1 \, \quad \, i_X \star i_n \star g = 0 \, ,
\end{equation}
we find
\begin{equation}
[ n \mb N\star n] \star g +  [n \mb M\star X] \star g = {\rm d}N+({\rm
  d}N\star i_n)\star n
\end{equation}
or
\begin{equation}
  i_n\star {\rm d} N\star n^{\mu}\star g_{\mu\nu}+ [n\mb M\star X]^{\mu}\star
 g_{\mu\nu}= \partial_{\nu}N+
 \partial_\gamma N\star n^\gamma \star n_\nu\,.
\end{equation}
We extract the tangential part by $\star$-multiplying both sides of the
equation by $g^{\nu\alpha}\star q^a_\alpha$ from the right
\begin{equation}\label{tan2}
[n \mb M\star X]^a = \partial_\nu N \star g^{\nu\alpha}\star q^a_\alpha \, ,
\end{equation}
and the normal part by $\star$-multiplying both sides of the equation by
$n^\nu$ from the right
\begin{equation}
\label{norm2}
-n^\rho\star \partial_\rho N + [n\mb M\star X]^{\mu}\star g_{\mu\nu}\star
n^{\nu}=0\,. 
\end{equation}
The commutator term is equal to
\begin{equation}
  [n\mb M\star X]^{\mu} \star g_{\mu\nu}\star n^{\nu} = n^{\rho}
  \star \partial_{\rho} (M\star X)^{\mu} \star g_{\mu\nu}\star
  n^{\nu}- (M\star
  X)^{\rho}\star (\partial_{\rho}n^{\mu}) \star g_{\mu\nu}\star n^{\nu}\,.
\end{equation}
In our Gaussian frame, $n^{\alpha}$ is normalized, geodesic, and has constant
components. The commutator is therefore zero and we have
\begin{equation}
n^\nu\star \partial_\nu N  = 0 \,.
\end{equation}
Since the components $n^{\nu}$ are constant, the $\star$-product does not
imply higher derivatives in this equation. Therefore, we still have a
well-posed initial-value problem for lapse $N$ and shift $M^a$.

Using a decomposition as in (\ref{vvGen}), we now obtain
\begin{equation} \label{HDAstar}
[(N_1,0) \mb (N_2,0)] = (0,  (N_1 \star \partial_b N_2 - N_2\star \partial_b
N_1 )  \star q^{ab}) \, . 
\end{equation}
For brackets involving tangential vector fields, we have
\begin{equation}\label{MM2}
[(0,M^a_1) \mb (0,M^a_2)] =\left( 0 , [M_1 \star X \mb M_2
  \star X]^\alpha \star q^a_\alpha \right)
\end{equation}
and
\begin{equation}\label{NM1}
[(N,0) \mb (0,M^a)]  = \left( -\mathcal{L}_{M\star X}\triangleright N ,0
\right)\,.
\end{equation}

Therefore, we are able to derive a well-defined HDA in our modified Gaussian
frame. It has the form of the classical version without any correction term
other than a generalization to Moyal space. This means that we find for the
$\star$-HDA the same form of the classical HDA but with the usual point
product replaced by the $\star$-product. Note, however, that the
$\star$-product implies higher time derivatives which affect the
interpretation of the HDA. We will comment on this implication in more detail
in Section~\ref{s:DefDef}.

We now have a possible candidate for a $\star$-HDA. According to
Ref.~\cite{szabo1}, for instance, once the deformation of infinitesimal
diffeomorphisms has been introduced, the action for gravity should be written
with the requirement of invariance under these $\star$-symmetries. In
particular, the deformed Einstein--Hilbert action should be formulated in
terms of star-products. However, in order to make sure that there is a
  fully covariant tensor calculus, we have to return to a discussion of the
  Leibniz rule. 

\subsection{Modified Leibniz rule}

The demonstration that an action for noncommutative gravity, such as the one
introduced in Ref.~\cite{aschieri1}, is covariant requires an application of
the Leibniz rule. In particular, inserting the Lie derivative $\delta_vL$ in
the Lagrangian {\em density} $L=E\star R$ in an action
\begin{equation}
S_\star = \int \, {\rm d}^4 x L =  \int \, {\rm d}^4x E\star R\,,
\end{equation}
where ${\rm d}^4x E$ is a suitably deformed measure and $R$ is the
$\star$-Ricci scalar, should result in a boundary term. (See, for instance,
\cite{aschieri1} for details and explicit expressions.)

Assuming the Leibniz rule, the infinitesimal variation of the Lagrangian
density under deformed diffeomorphisms would be given by
\begin{eqnarray} \label{Lie1}
  \delta_v L &=& \mathcal{L}_v \triangleright (E \star R) = (\mathcal{L}_v
  \triangleright E) \star R+ E\star \mathcal{L}_v \triangleright R\\
  &=& (v^\rho \star \partial_\rho E+ \partial_\rho v^\rho \star E) \star R + E
  \star v^\rho \star \partial_\rho R \\
  &=& \partial_{\rho} (v^{\rho} \star E \star R) +E \star v^{\rho}
  \star \partial_{\rho}   R - v^{\rho} \star E \star \partial_{\rho} R \,,
\end{eqnarray}
which differs from a total derivative by the non-zero star commutator $(E
\star v^{\rho}-v^{\rho}\star E)\star \partial_{\rho}R$. However, foregoing the
Leibniz rule at this point and applying the Lie derivative directly to
the density $E\star R$ does give us a total derivative:
\begin{eqnarray}
 \mathcal{L}_v \triangleright (E \star R) &=& 
 v^{\rho}\star\partial_{\rho}(E\star R)+ (\partial_{\rho}v^{\rho}\star)
 (E\star R)\\
&=& \partial_{\rho}(v^{\rho}\star E\star R)\,.
\end{eqnarray}
The action would then be invariant but the Lie derivative does not agree with
(\ref{Lie1}).

We therefore have to refine our notion of deformed diffeomorphisms, in
contrast to the situation in noncommutative field theories
\cite{szabo1,szabo2}, for which there are $\star$-actions invariant under both
twisted ${\rm U}(N)$ transformations with non-trivial coproducts and deformed
${\rm U}(N)$ transformations with standard Leibniz rule \cite{douglas1}. The
main reason why we tried to define $\star$-diffeomorphisms with trivial
co-multiplication was the desire to mimic what happens in noncommutative
quantum field theories, but we now see that there is a pronounced difference
between noncommutative gravity and other noncommutative systems at a
fundamental level.

In our example of a density times the Ricci scalar, the defect in the Leibniz
rule was given by a star commutator of components. We can therefore try to
modify the Leibniz rule by rearranging different factors.  We now define
\begin{equation}\label{defL}
\mathcal{L}_v \triangleright (u \star w) :=  (\mathcal{L}_v \triangleright u)
\star w + \overline{R}(u) \star (\mathcal{L}_{\overline{R}(v)}
\triangleright w) \, , 
\end{equation}
where $\overline{R}$ is defined in (\ref{Rmat}). Together with this deformed 
Leibniz rule, we also change the ordering in the action of 
$\star$-diffeomorphisms on vectors to obtain the new Lie derivative
\begin{equation}
\mathcal{L}_v \triangleright u^\mu := v^\rho \star \partial_\rho u^\mu
-  \partial_\rho v^\mu \star u^\rho \, . 
\end{equation}

Now we can prove that $u^\mu \star u_\mu$ transforms as a
scalar under deformed diffeomorphisms: We have 
\begin{equation}
\begin{split}
(\mathcal{L}_v \triangleright u^\mu )
\star u_\mu + \overline{R}(u^\mu) \star (\mathcal{L}_{\overline{R}(v)}
\triangleright u_\mu) \\ = (v^\rho 
\star \partial_\rho u^\mu -  \partial_\rho v^\mu \star u^\rho) \star u_\mu +
\overline{R}(u^\mu) \star (\overline{R}(v^\rho) \star \partial_\rho u_\mu +
\overline{R}(\partial_\mu v^\rho ) \star u_\rho) \\ = v^\rho
\star \partial_\rho u^\mu \star u_\mu -  \partial_\rho v^\mu \star u^\rho
\star u_\mu + v^\rho \star u^\mu \star \partial_\rho u_\mu + \partial_\mu
v^\rho \star u^\mu \star u_\rho \, .
\end{split}
\end{equation}
The second and the fourth terms in the last line cancel out, and we have
\begin{equation}
 (\mathcal{L}_v \triangleright u^\mu )
\star u_\mu + \overline{R}(u^\mu) \star (\mathcal{L}_{\overline{R}(v)}
\triangleright u_\mu) =   v^\rho \star \partial_\rho u^\mu \star
u_\mu  + v^\rho \star u^\mu \star \partial_\rho u_\mu = \mathcal{L}_v
\triangleright (u^\mu \star u_\mu ) \, . 
\end{equation}

In order to prove that the new Leibniz rule implies a consistent extension of
the deformed Lie derivative to tensors, we start with the $\star$ product of
two vector fields, $u^\mu_1 \star u^\nu_2$:
\begin{eqnarray}
 \mathcal{L}_v \triangleright (u^\mu_1 \star u^\nu_2) =  (\mathcal{L}_v
 \triangleright u^\mu_1 ) 
\star u^\nu_2 + \overline{R}(u^\mu_1)  \star (\mathcal{L}_{\overline{R}(v)}
\triangleright u^\nu_2) \nonumber\\
=  v^\rho\star \partial_\rho u^\mu_1 \star u^\nu_2  - \partial_\rho v^\mu
\star u^\rho_1 \star u^\nu_2 + v^\rho\star u^\mu_1 \star \partial_\rho u^\nu_2
- \partial_\rho v^\nu \star u^\mu_1 \star u^\rho_2 \nonumber \\ 
= v^\rho \star \partial_\rho (u^\mu_1 \star u^\nu_2) - \partial_\rho v^\mu
\star u^\rho_1 \star u^\nu_2 - \partial_\rho v^\nu \star u^\mu_1 \star
u^\rho_2 =  \mathcal{L}_v \triangleright  T^{\mu\nu} 
\end{eqnarray}
with the contravariant 2-tensor $T^{\mu\nu} := u^\mu_1 \star u^\nu_2$. By
induction, the claim then follows for arbitrary tensors:
\begin{eqnarray}
 \mathcal{L}_v \triangleright (u^{\mu_1}_1 \star u^{\mu_2}_2\star \dots \star
 u^{\mu_n}_n \star w^1_{\nu_1}\star \dots \star w^n_{\nu_n}) \nonumber \\
= ( \mathcal{L}_v \triangleright u^{\mu_1}_1 )\star (u^{\mu_2}_2\star \dots
\star u^{\mu_n}_n \star w^1_{\nu_1}\star \dots \star w^n_{\nu_n}) \nonumber\\
+ \overline{R}_1(u^{\mu_1}_1
)\star(\mathcal{L}_{\overline{R}_1(v)}\triangleright  (u^{\mu_2}_2\star \dots
\star u^{\mu_n}_n \star w^1_{\nu_1}\star \dots \star w^n_{\nu_n})) \nonumber\\
= (v^\rho \star \partial_\rho  u^{\mu_1}_1 )\star (u^{\mu_2}_2\star \dots
\star u^{\mu_n}_n \star w^1_{\nu_1}\star \dots \star w^n_{\nu_n})\nonumber \\
- (\partial_\rho v^{\mu_1}\star u^\rho_1) \star (u^{\mu_2}_2\star \dots \star
u^{\mu_n}_n \star w^1_{\nu_1}\star \dots \star w^n_{\nu_n})\nonumber \\
+v^\rho \star u^{\mu_1}\star \partial_\rho u^{\mu_2}_2 \star u^{\mu_3}_3\star
\dots \star u^{\mu_n}_n\star w^1_{\nu_1}\star \dots \star
w^n_{\nu_n}\nonumber\\ 
- \partial_\rho v^{\mu_2}\star u^{\mu_1}_1\star u^\rho_2 \star
u^{\mu_3}_3\star \dots \star u^{\mu_n}_n\star w^1_{\nu_1}\star \dots \star
w^n_{\nu_n} \nonumber\\
+ \overline{R}_1(u^{\mu_1}_1 )\star \overline{R}_2(u^{\mu_2}_2)\star(
\mathcal{L}_{\overline{R}_2\overline{R}_1(v)}\triangleright ( u^{\mu_3}_3\star
\dots \star u^{\mu_n}_n \star w^1_{\nu_1}\star \dots \star
w^n_{\nu_n}))\nonumber \\ 
=\cdots= \mathcal{L}_v  \triangleright (  T^{\mu_1\mu_2 \dots
  \mu_n}_{\nu_1\nu_2\dots \nu_n})
\end{eqnarray}
with $T^{\mu_1\mu_2 \dots \mu_n}_{\nu_1\nu_2\dots \nu_n} := u^{\mu_1}_1 \star
u^{\mu_2}_2\star \dots \star u^{\mu_n}_n \star w^1_{\nu_1}\star \dots \star
w^n_{\nu_n}$.


\subsection{Deformed diffeomorphisms with deformed Leibniz rule}
\label{s:DefDef}

We have clarified the reason why the Leibniz rule has to be modified when we
adopt a noncommutative multiplication rule, and provided a new definition to
resolve the problem. With this result, we can now focus on the derivation of
the hypersurface-deformation brackets for deformed diffeomorphisms with
deformed Leibniz rule as in Eq.~\eqref{defL}.

Combining the lessons from our previous derivation with the standard Leibniz
rule as well as the new Lie derivative, we now introduce a modified Gaussian
condition by requiring
\begin{equation}
\overline{R}(n^\mu) \star (\mathcal{L}_{\overline{R}(v)} \triangleright g_{\mu
  \nu }) = -\partial_\rho (v^\rho \star n^\nu \star g_{\mu\rho})\star n^\gamma
\star n^\rho \star g_{\gamma\nu}  
\end{equation}
for space-time vector fields $v$.  Using the modified Leibniz rule we can
rewrite this equation as
\begin{equation}
\mathcal{L}_v \triangleright(n^{\mu}\star g_{\mu\nu} ) -( \mathcal{L}_v
\triangleright n^\mu)\star g_{\mu \nu} =  -\partial_\rho (v^\rho \star n^\nu
\star g_{\mu\rho})\star n^\gamma \star n^\rho \star g_{\gamma\nu} \, ,  
\end{equation}
and thanks to the Cartan identity, obtain
\begin{equation}
\partial_\nu ( v^\rho \star n^\mu \star g_{\mu\rho}) +  v^\rho \star
({\rm d}n)_{\rho\nu} - [v , n]^\mu_\star \star g_{\mu\nu} =  -\partial_\rho
(v^\rho \star n^\nu \star g_{\mu\rho})\star n^\gamma \star n^\rho \star
g_{\gamma\nu} \, .  
\end{equation}
Here $({\rm d}n)_{\rho\nu} \equiv \partial_\rho (n^\mu \star g_{\mu\nu})
- \partial_\nu ( n^\mu \star g_{\mu \rho}) $ vanishes as before.
Decomposing $v^\mu = N\star n^\mu +
M^a\star X^\mu_a$, we find
\begin{equation} \label{partialNDef}
-\partial_\nu N - [N\star n , n]^\mu_\star \star g_{\mu\nu} - [M\star X ,
n]^\mu_\star \star g_{\mu \nu} =  \partial_\rho N  \star n^\gamma \star n^\rho
\star g_{\gamma\nu}  \, .  
\end{equation}

Projection implies the normal part
\begin{eqnarray*}
&&-\partial_\nu N\star g^{\nu\alpha}\star n^\beta \star g_{\alpha\beta} -
[N\star n , n]^\alpha_\star \star  n^\beta \star g_{\alpha\beta}  - [M\star X, 
n]^\alpha_\star \star n^\beta \star g_{\alpha\beta} \\
&=&  \partial_\rho N  \star n^\gamma \star n^\rho \star g_{\gamma\nu} \star
g^{\nu\alpha}\star n^\beta \star g_{\alpha\beta}  \, ,  
\end{eqnarray*}
or
\begin{eqnarray*}
&&-\partial_\nu N\star n^\nu - N\star n^\rho \star \partial_\rho n^\alpha
\star  n^\beta \star g_{\alpha\beta} + \partial_\rho (N \star n^\alpha) \star
n^\rho \star  n^\beta \star g_{\alpha\beta}\\
&&  - [M\star X ,n]^\alpha_\star
\star n^\beta \star g_{\alpha\beta}\\  &=&  \partial_\rho N  \star n^\alpha
\star n^\rho \star n^\beta \star g_{\alpha\beta}  \, .
\end{eqnarray*}
We now use $n^\rho \star \partial_\rho n^\mu = 0$, cancel
out $\partial_\rho N \star n^\alpha \star n^\rho \star n^\beta \star
g_{\alpha\beta}$, and obtain
\begin{equation}
-\partial_\nu N\star n^\nu = [M\star X ,n]^\alpha_\star \star n^\beta \star
g_{\alpha\beta} \, .  
\end{equation}
The commutator on the right is equal to
\begin{equation}
 [M\star X,n]^{\alpha}_{\star} \star n^{\beta}\star g_{\alpha\beta} = (M\star
 X)^{\gamma}\star (\partial_{\gamma}n^{\alpha}) \star n^{\beta}\star
 g_{\alpha\beta}- n^{\gamma} \star \partial_{\gamma} (M\star X)^{\alpha} \star
 n^{\beta}\star g_{\alpha\beta}\,.
\end{equation}
If we now use the properties of our Gaussian frame, in particular that
$n^{\alpha}$ is normalized, geodesic, and has constant components, the
commutator is zero and we arrive at
\begin{equation} \label{NDef}
-\partial_\nu N\star n^\nu = 0 \,.
\end{equation}
The tangential part of (\ref{partialNDef}) is
\begin{eqnarray}
&&-\partial_\nu N\star g^{\nu\alpha}\star q_{\alpha b}  - [M\star X
,n]^\alpha_\star \star q_{\alpha b} - [N\star n , n]^\alpha_\star \star
q_{\alpha b}\\
&=&  -\partial_b N  - [M\star X ,n]^\alpha_\star \star q_{\alpha b}  - N\star
n^\rho \star \partial_\rho n^\alpha \star q_{\alpha b} - \partial_\rho N \star
n^\alpha \star n^\rho \star q_{\alpha b}\\
&=& \partial_\rho N \star n^\alpha \star n^\rho \star q_{\alpha b}
\, ,  
\end{eqnarray}
which is equivalent to
\begin{equation} \label{MDef}
 [M\star X ,n]^a_\star = - \partial_b N \star q^{ab} \, . 
\end{equation}
As before, the equations for lapse and shift provide a well-posed
initial-value problem.

We can now compute the bracket 
\begin{eqnarray}
[v_1 , v_2 ]^\mu_\star &=& [N_1 \star n , N_2 \star n]^\mu_\star +[N_1 \star n ,
M_2 \star X ]^\mu_\star \\
&&+ [M_1\star X , N_2 \star n ]^\mu_\star + [M_1 \star
X , M_2 \star X]^\mu_\star  \nonumber\\
 &=& N_1 \star n^\rho \star \partial_\rho ( N_2 \star n^\mu ) 
- \partial_\rho (N_1\star n^ \mu) \star N_2 \star n^\rho \nonumber \\ 
&&+ N_1 \star
n^\rho \star \partial_\rho (M_2 \star X^\mu) - \partial_\rho (N_1 \star n^\mu ) \star M_2
\star X^\rho  + M_1 \star X^\rho \star \partial_\rho (N_2 \star n^\mu) \nonumber\\
&&- \partial_\rho
(M_1 \star X^\mu) \star N_2 \star n^\rho  + [M_1 \star X , M_2 \star X]^b_\star
\star X^a_b \, .
\end{eqnarray}
Choosing Gaussian vector fields with either zero lapse $N$ or shift $M^a$
functions we can decompose the above brackets as a set of three distinct
commutators $[(0, M_1) , (0, M_2) ]_\star$, $ [( 0, M_1) , (N_2,0) ]_\star$ and
$[(N_1, 0) , (N_2, 0) ]_\star$. If both lapse functions are zero, we find
\begin{equation}
[(0, M_1) , (0, M_2) ]_\star =  ( 0 , [M_1 \star X , M_2 \star X]^a_\star) \, .
\end{equation}
For both shift vector fields equal to zero, we obtain
\begin{equation} \label{NNstar}
[(N_1, 0) , (N_2, 0) ]_\star = ( 0, N_1 \star q^{a b}\star \partial_b N_2
- \partial_b N_1  \star N_2 \star q^{a  b})  \, . 
\end{equation}
The remaining bracket reads
\begin{equation}
 [( 0, M) , (N,0) ]_\star  = (\mathcal{L}_{M\star X} \triangleright N , 0)
 \, .  
\end{equation}

It is perhaps surprising that the overall structure of the bracket between $N$
and $M^a$ is preserved despite the noncommutativity of coordinates. In this
regard, one can note that the $\star$-Lie bracket between two tangential
deformations still gives us a tangential hypersurface deformation, the one
involving a normal and a tangential deformations gives a normal displacement,
and the bracket between two normal deformations results in a spatial
shift. The only type of modifications that appear with respect to the standard
hypersurface brackets are higher derivative terms. Those terms are implicit in
the above expressions, but it is clear that such terms appear as soon as we
expand the Moyal star product by powers of $\theta$. 

Although the brackets bear a formal resemblance with the classical ones,
their detailed form is markedly different. The main reason is the
non-locality of the $\star$-product, which includes higher derivatives in
{\em space-time}. In the noncommutative HDA brackets as written, we
therefore have time derivatives of $N$, $M^a$ and the inverse spatial metric
$q^{ab}$, which, unlike those of the constant $n^{\mu}$, are in general
non-zero. Since the brackets cannot contain space-time data, we should
interpret the $\star$-products in them as follows: Working in the Gaussian
frame, time derivatives of $N$ and $M^a$ can be replaced by spatial
derivatives using the equations (\ref{NDef}) and (\ref{MDef}). Any
first-order time derivative of $q_{ab}$ can be expressed as a linear
combination of extrinsic-curvature components $K_{ab}$, while higher-order
time derivatives of $q_{ab}$ are related to higher-order momenta in the
Ostrogradsky treatment of a canonical higher-derivative theory. Without a
specific noncommutative action, we cannot write these terms explicitly, but
rather leave the brackets in the form (\ref{NNstar}) with implicit
higher-derivative terms.

We conclude that the base manifold of the noncommutative HDA should
contain not only the spatial metric but the entire phase space of
a higher-derivative metric theory. The presence of extrinsic curvature among
these variables is reminiscent of holonomy modifications in models of loop
quantum gravity, but the explicit dependence is, in general, different (see e.g.
\cite{holocorr1,holocorr2}).


\section{Conclusions}
\label{s:Concl}

We have studied infinitesimal diffeomorphisms on noncommutative manifolds
equipped with a non-standard multiplication rule in terms of
$\star$-products. Previous studies on noncommutative formulations of gravity
(in particular \cite{aschieri1,aschieri2}) succeeded in twisting the group of
4-dimensional diffeomorphisms, thereby achieving a deformation of GR
symmetries in the sense of Drinfeld twists \cite{drin,DT}. Nonetheless, as
already pointed out in the literature and further stressed in this work, it
remains unclear whether diffeomorphisms on noncommutative spaces should be
introduced by means of twisting or explicitly deforming their action, as it is
the case for $\star$-gauge transformations in noncommutative extensions of
quantum field theories. 

The study of the algebra of hypersurface deformations, generating
diffeomorphisms if we make a $3+1$ splitting of the 4-manifold, can provide
additional insights into general covariance in noncommutative gravity as well
as on the counting of physical degrees of freedom of the theory.  Our analysis
is one of only a few in the context of canonical formulations of
noncommutative gravity. In addition to shedding some light on long-standing
questions in noncommutative gravity, it might also help in making contact with
other recently proposed modifications of the HDA
\cite{holocorr1,holocorr2,holocorr3,perez,anomlqc,vara,tomlin,frahda}. One
possible point of contact is the presence of extrinsic curvature as one of the
coordinates on the base manifold of a noncommutative HDA.

By using a recently developed approach to the derivation of the HDA
\cite{wein}, we have shown a constructive method to derive the brackets
between spatial and time components of Gaussian vector fields when functions
and tensors are multiplied with a noncommutative $\star$-product. This
application is conceptually different from the derivation in classical general
relativity given in \cite{wein} because we cannot take for granted that there
is a covariant theory with a well-defined HDA on noncommutative manifolds. We
therefore had to demonstrate that the frame of a Gaussian system, used in
\cite{wein}, can be suitably generalized to specific types of
noncommutativity. After doing this, we derive well-defined HDAs, which implies
that there are infinitesimal space-time transformations that allow us to
change the frame. In this sense, we have demonstrated the covariance of such
theories, even though we did not use an explicit action principle.  

In particular, we have studied both the HDA encoding twisted diffeomorphisms
and the deformations of the HDA produced by what we call deformed or
$\star$-diffeomorphisms. In the former case, we have found, not surprisingly,
that the brackets are unmodified compared with the classical algebra of GR
gravitational constraints. This result confirms some of the previous
statements that appeared in the literature on twisted gravity
\cite{aschieri1}.

In the analysis of the latter case --- deformed diffeomorphisms --- we did not
have any guidance from established results. Thus, building on the analogy with
$\star$-${\rm U}(1)$ (or in general $\star$-${\rm U}(N)$) gauge theories, we
first defined deformed diffeomorphisms with a suitably deformed action on
single fields but retaining the Leibniz rule in their action on the
$\star$-product of two or more functions. We were able to overcome the
technical challenges represented by the correction terms to the HDA brackets,
but eventually recognized a major obstacle to the implementation of a
consistent noncommutative differential calculus where diffeomorphism
transformations have a trivial coalgebra. This forced us to deform the
coproducts of $\star$-diffeomorphisms. As a result, we have reached a
meaningful deformation of the HDA for deformed diffeomorphisms without
pathological correction terms and with a consistent differential calculus
suitably adapted to $\star$-products. 

While formally similar to the classical HDA, noncommutative HDAs based on
deformed diffeomorphisms show crucial differences in their structure owing to
non-locality (in particular in time) of $\star$-products. We hope that this result may serve
as a base for an alternative formulation of noncommutative gravity in terms of
the deformed diffeomorphisms put forward here, instead of relying on the
symmetry principle of twisting as done so far. The path we followed here
provides a simplified way to get insight into how general covariance might be
affected by $\star$-products or other possible deformations.

For twisted diffeomorphisms we have also been able to discuss
straightforwardly the flat-spacetime (or Minkowski) limit since we had no
deformations of the HDA.  On the contrary, the study of the Minkowski regime
of the deformed HDA encoding $\star$-diffeomorphisms remains an open challenge
which should be of particular interest both from the perspective of relating
$\star$-product corrections to the non-linear Poincar\'e transformations of
noncommutative spacetimes \cite{majrue,lukrue} and also to have a better
understanding of what general modifications of the HDA should affect the
Poincar\'e algebra.

\paragraph*{Acknowledgments.}

This work was supported in part by NSF grant PHY-1607414. MR thanks the
Institute for Gravitation and the Cosmos and Penn State University for the
hospitality during the elaboration of this project. The work of MR is
supported by a mobility grant awarded by Sapienza University of Rome and
MIUR. The contribution of MR is based upon work from COST Action MP1405
QSPACE, supported by COST (European Cooperation in Science and Technology).

\begin{appendix}

\section{Modified Gaussian condition: Classical case}
\label{a:ModGauss}

In this appendix, we show that a suitable modification \eqref{defgf} of the
Gaussian system still results in the usual classical HDA.  We begin with
\begin{equation}
n^\mu \mathcal{L}_v g_{\mu\nu} = n^\rho \partial_\rho (g_{\delta\gamma}
n^\gamma v^\delta) n_\nu \, . 
\end{equation}
and write the Lie derivative explicitly:
\begin{equation}
n^\mu v^\sigma \partial_\sigma g_{\mu\nu} + n^\mu g_{\mu\sigma}\partial_\nu
v^\sigma + n^\mu g_{\sigma\nu}\partial_\mu v^\sigma =  n^\rho \partial_\rho
(g_{\delta\gamma} n^\gamma v^\delta) n_\nu \, , 
\end{equation}
or, equivalently,
\begin{equation*}
\partial_\nu (g_{\mu\sigma}n^\mu v^\sigma)-v^\sigma\partial_\nu (g_{\mu\sigma}
n^\mu) + v^\sigma  \partial_\sigma( g_{\mu\nu} n^\mu) -v^\sigma
g_{\mu\nu} \partial_\sigma n^\mu + n^\mu g_{\sigma\nu}\partial_\mu v^\sigma =
n^\rho \partial_\rho (g_{\delta\gamma} n^\gamma v^\delta) n_\nu \, . 
\end{equation*}
Using $({\rm d}n)_{\sigma\nu} := \partial_\sigma( g_{\mu\nu}
n^\mu) - \partial_\nu (g_{\mu\sigma} n^\mu)$ and $[n,v]^\mu =
n^\rho \partial_\rho v^\mu - v^\rho \partial_\rho n^\mu$, we obtain
\begin{equation*}
\partial_\nu (g_{\mu\sigma}n^\mu v^\sigma) + v^\sigma ({\rm d}n)_{\sigma\nu} +
[n,v]^\mu g_{\mu\nu} =  n^\rho \partial_\rho (g_{\delta\gamma} n^\gamma
v^\delta) n_\nu  \, .
\end{equation*}
If we choose the metric such that
\begin{equation} \label{Gauss}
{\rm d}s^2 = -{\rm d}t^2 + q_{ab} {\rm d}x^a {\rm d}x^b \, ,
\end{equation}
we have ${\rm d}n = {\rm d}^2 t = 0$. Although our Gaussian condition has
  been modified, the metric (\ref{Gauss}) is consistent with the gauge choice
  as shown by the final result, in particular Eq.~(\ref{normal}). Moreover,
\begin{equation}
\partial_\nu (g_{\mu\sigma}n^\mu v^\sigma) + [n,v]^\mu g_{\mu\nu} =
n^\rho \partial_\rho (g_{\delta\gamma}  n^\gamma v^\delta) n_\nu \, .
\end{equation}
Decomposing $v$ as $v^\mu = Nn^\mu + M^\mu$ we obtain
\begin{equation}
-\partial_\nu N + [n,M]^\mu g_{\mu\nu} = -n^\rho \partial_\rho N n_\nu \, .
\end{equation}

Let us now find the normal and tangential components of the above
equality. For the normal we have
\begin{equation}
-n^\nu \partial_\nu N + [n,M]^\nu n_\nu = n^\rho \partial_\rho N \, ,
\end{equation}
and thus
\begin{equation}
\label{normal}
n^\nu \partial_\nu N = 0 
\end{equation}
because $[n,M]$ does not have a normal component.  For the tangential part, we
obtain
\begin{equation}
\label{tangential}
[n,M]^a = q^{ab}\partial_b N  \, .
\end{equation}

We are now ready to compute the bracket between two vector fields:
\begin{equation*}
\begin{split}
[v_1, v_2] ^\mu = (N_1 \mathcal{L}_n N_2 - N_2 \mathcal{L}_n N_1 +
\mathcal{L}_{M_1} N_2 - \mathcal{L}_{M_2}N_1)n^\mu\\ - N_2 [n,M_1] + N_1 [n,
M_2] + [M_1, M_2]^\mu \\ = (N_1 n^\rho \partial_\rho N_2 - N_2
n^\rho \partial_\rho N_1 + M^b_1\partial_b N_2 - M^b_2 \partial_b N_1 )
n^\mu\\ +N_1 q^{ab}\partial_b N_2 - N_2 q^{ab}\partial_b N_1 +  [M_1, M_2]^a
\\ = ( M^b_1\partial_b N_2 - M^b_2 \partial_b N_1 ) n^\mu +
q^{ab}(N_1 \partial_b N_2 - N_2\partial_b N_1) \, , 
\end{split}
\end{equation*}
where we used Eqs.~\eqref{normal} and \eqref{tangential}. Finally, we can
extract the normal and tangential components of the brackets:
\begin{eqnarray}
&&[(0,M_1), (0,M_2)] = (0,   \mathcal{L}_{M_1}M_2) \\
&&[(N_1,0), (0, M_2)] = -(\mathcal{L}_{M_2} N_1, 0) \\
&&[(N_1,0), (N_2,0)] = (0, q^{ab}(N_1 \partial_b N_2 - N_2\partial_b N_1)) \, .
\end{eqnarray}
These are the brackets of Dirac's hypersurface-deformation
algebroid.

\end{appendix}

\end{document}